\documentclass[prx,
 reprint,
 superscriptaddress,
 amsmath,amssymb,
 aps,
]{revtex4-2}

\usepackage{graphicx}
\usepackage{dcolumn}
\usepackage{bm}
\usepackage{physics} 
\usepackage{pgffor}

\usepackage{xr}

\usepackage{mathrsfs}
\usepackage[normalem]{ulem}

\usepackage{xcolor}
\usepackage{soul}

\usepackage{hyperref}
\hypersetup{
    colorlinks,
    linkcolor={black},
    citecolor={black},
    urlcolor={black}
}

\begin{document}

\preprint{APS/123-QED}

\title{\textbf{Second order N-Electron Valence state Perturbation Theory for polaritonic systems}}
\author{Edoardo Ascani}
\affiliation{Dipartimento di Chimica, Biologia e Biotecnologie, Universit{\`a} degli Studi di Perugia, 06123, Perugia, Italy}

\author{Riccardo Alessandro}
\email{riccardo.alessandro@dottorandi.unipg.it}
\affiliation{Dipartimento di Chimica, Biologia e Biotecnologie, Universit{\`a} degli Studi di Perugia, 06123, Perugia, Italy}

\author{Celestino Angeli}
\affiliation{Dipartimento di Scienze Chimiche, Farmaceutiche ed Agrarie, Universit{\`a} degli Studi di Ferrara, Ferrara, Italy}

\author{Enrico Ronca}
\email{enrico.ronca@unipg.it}
\affiliation{Dipartimento di Chimica, Biologia e Biotecnologie, Universit{\`a} degli Studi di Perugia, 06123, Perugia, Italy}

\begin{abstract}
Molecules strongly coupled to optical cavities often show multi-configurational character, which the cavity can enhance by mixing electronic states. Describing them accurately requires a complete treatment of correlation between all the involved quantum particles. The available QED \textit{ab-initio} methods do not treat it in its completeness except for QED-FCI. We extend second-order N-electron valence state perturbation theory (NEVPT2) to polaritonic chemistry, in its state-specific (QED-NEVPT2) and quasi degenerate (QED-QD-NEVPT2) forms. Here, the correlation due to the interaction of more degenerate configurations comes from a QED-CASSCF reference, while that associated with instantaneous interactions between the particles is added  by second-order perturbation theory. Application of the methods to ammonia, ethylene and formaldimine revealed that this approach represents the perfect alternative to the well established QED-Coupled Cluster approach when the molecular system is described by more degenerate configurations. The approaches developed in this work have the potential to become state-of-the-art methodologies for the description of reactivity and photochemistry in polaritonic systems.
\end{abstract}

\maketitle

\section{\label{sec:intro} Introduction}
Polaritonic chemistry is attracting the attention of the physics and chemistry communities since a decade~\cite{ebbesen2023introduction, mandal2023theoretical}. 
Many experiments demonstrated that strong coupling between matter and confined electromagnetic modes can modify the molecular electronic-state structure~\cite{eizner2019inverting}, photochemical pathways~\cite{hutchison2012modifying}, and even ground-state reaction rates~\cite{ebbesen2016hybrid, patrahau2024direct, thomas2016ground, thomas2019tilting}.

\begin{figure*}[ht!]
    \centering
    \includegraphics[width=\linewidth]{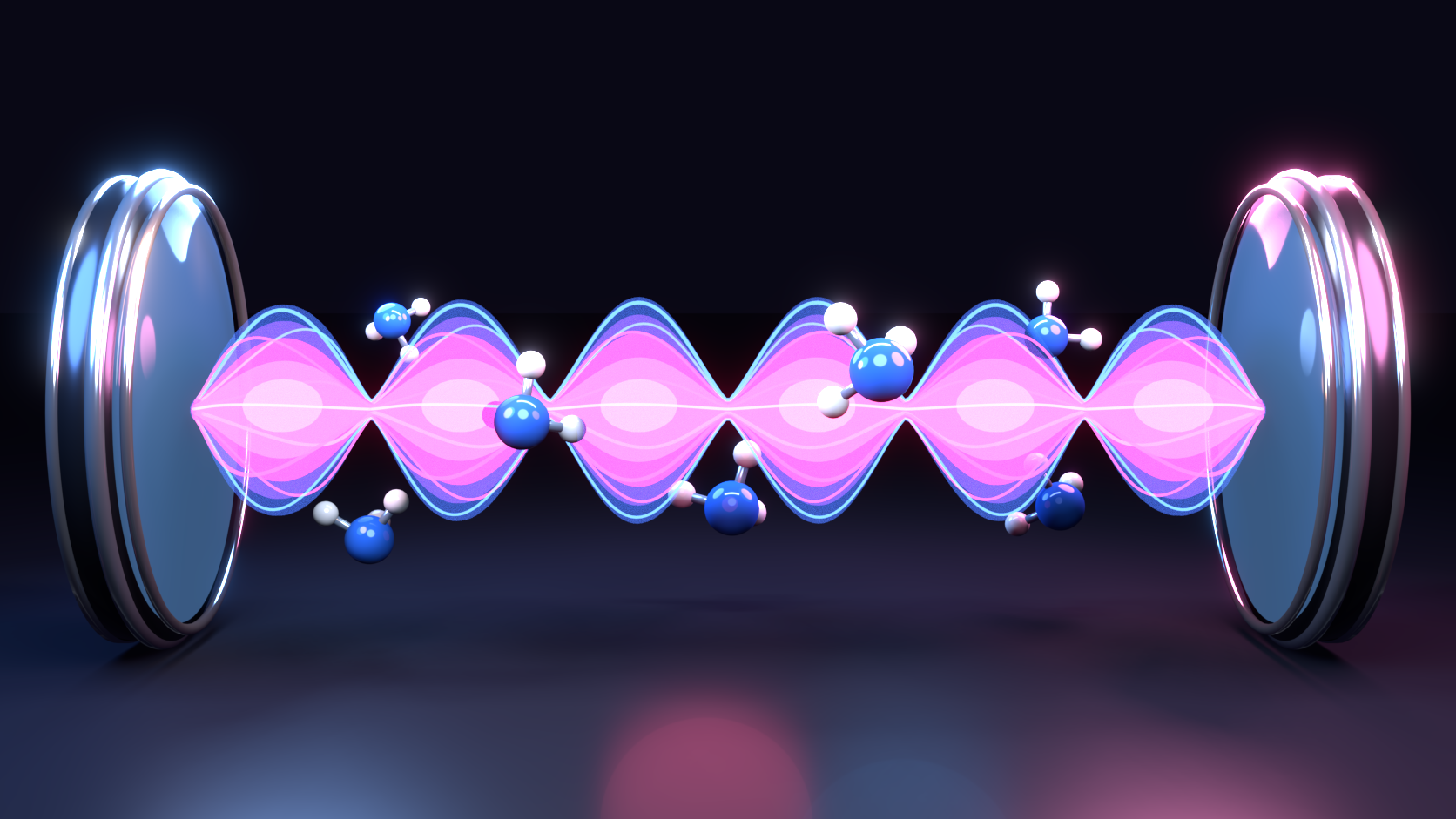}
    \caption{Molecules coupled to a quantum field in an optical cavity}
    \label{fig:Cartoon}
\end{figure*}

This regime is usually obtained in optical cavities (Fig.~\ref{fig:Cartoon}).
In these conditions, the matter and the field states couple to generate new hybrid states called "\textit{polaritons}" characterized by properties that, in some cases, can be also very different from the bare matter ones.
Depending on the frequency range, polaritons can be obtained by hybridization with electronic (Electronic Strong Coupling -- ESC) or vibrational (Vibrational Strong Coupling -- VSC) molecular states. 
In the first case (ESC) significant modifications of the excited states potential energy surfaces can be induced with consequent major effects on the photochemistry of the molecular system~\cite{ebbesen2016hybrid, orgiu2015conductivity, martinez2018polariton, delpo2020polariton, polak2020manipulating}.
If the field is in resonance with vibrational excitations (VSC) instead ground-state reaction rates modifications can in principle be induced~\cite{thomas2016ground, thomas2019tilting, lather2019cavity, sau2021modifying}.

Despite many experimental and theoretical groups have dedicated years of research to clarify the fundamental physics governing this phenomenon, in particular for VSC, a strong debate is still waging on the actual relevance of the field effects on the molecular reactivity~\cite{lather2019cavity, vergauwe2019modification, vurgaftman2020negligible, nagarajan2021chemistry, imperatore2021reproducibility, wiesehan2021negligible, climent2020sn2, ahn2023modification, hirai2023molecular, thomas2024selection, michon2024impact}.

In this context, emerged the urgency for new theoretical methodologies to describe the system (molecules, field, solvent, cavity dissipation, etc.) in its totality in an accurate manner.
This is clearly a major challenge considering that all the systems components require, in principle, a correlated quantum treatment.
Focusing on the strong interaction between a molecular system and the cavity field, in the past years several \textit{ab-initio} approaches~\cite{ruggenthaler2014quantum, ruggenthaler2018quantum, flick2018ab, haugland2020coupled, mordovina2020polaritonic, deprince2021cavity, pavosevic2021polaritonic, pavovsevic2022wavefunction, riso2022molecular, liebenthal2022equation, riso2023strong, foley2023ab, matousek2024polaritonic, el2024toward, konecny2025relativistic, castagnola2025strong, thiam2025comprehensive, el2025strong, alessandro2025complete, vu2025modeling, weight2025auxiliary, weber2025phaseless, zhang2026scalable, yuwono2026dynamic, deprince2026cavity} have been developed to perform simulations at different level of accuracy.

The \textit{ab-initio} simulation of molecular polaritons requires the extension of electronic structure theory methods to Quantum Electrodynamics (QED) Hamiltonians accounting for the simultaneous quantum nature of both the molecule and the field~\cite{bauman2025perspective}.

Between these approaches, QED-Coupled Cluster (QED-CC)~\cite{haugland2020coupled, mordovina2020polaritonic, pavosevic2021polaritonic, deprince2021cavity, liebenthal2022equation, monzel2024diagrams, monzel2026polaritonic, yuwono2026dynamic} -- the natural extension of coupled cluster theory\cite{bartlett2007coupled, shavitt2009many, helgaker2013molecular} to interacting electron photon systems -- took the scene for its high level of accuracy in the description of both the electron-electron and electron photon correlation.
Recently an extension of the method based on an \textit{ab-initio} treatment of one molecule coupled to a model treatment of the others has been also proposed to include collective effects~\cite{castagnola2025realistic}. 
This approach has been developed to face the high computational cost of CC in simulating a macroscopic number of molecules. 
Its application allows for accurate electronic structure calculation without the need to using unphysical values for the electron-photon coupling.

Despite being very accurate in describing \textit{dynamical correlation} -- instantaneous electron-electron interaction during their motion -- CC dramatically fails when degeneracies between different states (i.e. conical intersections, etc.) or between equivalent electronic configurations (i.e. complexes containing $d$ or $f$ partially occupied orbitals) come into play~\cite{szalay2012multiconfiguration, lischka2018multireference, choudhury2024understanding}.
In these cases multi-reference methodologies that properly account for \textit{static correlation} -- electron interaction associated with degeneracy between more configurations -- need to be applied. 
It is important to highlight that the coupling to the photons can introduce additional mixing between electronic states enhancing, in some cases, the multi-configurational character of the system.   

Multiconfiguration self-consistent field (MCSCF) approaches are based on a multideterminantal representation of the wavefunction and have been historically adopted for the description of multireference systems~\cite{hinze1973mc, shepard1987multiconfiguration, helgaker2013molecular}.
Complete active space (CAS) approaches are a category of MCSCF methods where the orbital space is divided into three subspaces: inactive, active, and virtual orbitals, and the determinants for the expansion are obtained from the distribution of a certain number of electrons within the active space~\cite{roos1980SCI_Dens, roos1980SCI_Fock, roos1987complete}.
Despite being very efficient in recovering the \textit{static correlation}, their abilities in including \textit{dynamical correlation} are very reduced.

This limitation is usually addressed by combining these approaches with other techniques able to account for the instantaneous interactions between the electrons~\cite{chattopadhyay2016state}.
One of the most used strategies to recover \textit{dynamic correlation} is perturbation theory. 
In this regard, while for single reference wavefunctions (within the Rayleigh-Schr{\"o}dinger formulation) the definition of a perturbation theory is straightforward, in the multiconfigurational case several choices of equivalent reference Hamiltonian can be chosen.
Among the possible MRPTs the second order Complete Active Space PT (CASPT2), based on a one electron zeroth order Hamiltonian, has emerged as a widely used methodology~\cite{andersson1990second, andersson1992second, battaglia2023multiconfigurational}.
Despite several limitations of CASPT2 have been identified and mitigated over time~\cite{lindh2020multi}, it suffers from a major drawback associated with the appearance of intruder states in the perturbation expansion~\cite{andersson1994cr2,camacho2009intruder}.
These unphysical states emerge from the one-electron nature of the zeroth order Hamiltonian which leads to an imbalance in the treatment of interactions between the active and outer spaces, producing vanishing energy denominators and resulting in discontinuities or unphysically large corrections in potential energy surfaces~\cite{Angeli2002}.
These limitations have been resolved with the formulation of the second-order N-electron valence state PT theory (NEVPT2)~\cite{Angeli2001, angeli2001n, Angeli2002, angeli2007new}.
This method developed by some of us, being based on a different reference Hamiltonian used to define the zeroth order bielectronic Hamiltonian (Dyall Hamiltonian~\cite{dyall1995choice}), is free from the emergence of intruder states. In this way it allows for a physical interpretation of the quantities involved in the perturbative treatment, such as the ionization potentials and the electron affinities as defined in the extended Koopmans' theorem~\cite{smith1975extension,Morrell1975}.

Standard single-state NEVPT2 works well for isolated states but cannot provide a proper description when two or more states become very close in energy (i.e., avoided crossings and conical intersections). This happens because each state is corrected independently without altering its zeroth-order configuration coefficients. This prevents the states from properly mixing and adapting under the perturbation, leading to artifacts such as spurious double crossings, as found with other MRPT2~\cite{spiegelmann1984use1,spiegelmann1984use}. 
These limitations are overcome by relying on its quasi-degenerate extension (QD-NEVPT2)~\cite{angeli2004quasidegenerate}, which incorporates multi-state interactions by diagonalizing an effective Hamiltonian matrix to prevent unphysical discontinuities.

In this paper we present the first extension of a MRPT to coupled electron-photon systems. 
It is based on the strongly contracted (SC-) formulation of NEVPT2 theory and its quasidegenerate version. 
The new theories take the name of QED-NEVPT2 and QED-QD-NEVPT2.
The proposed methodologies have been applied to investigate field-induced effects on a set of challenging multiconfigurational molecular systems.

The paper is structured as follows:
in Sec.~\ref{sec:electronic}, a brief summary of the (SC-)NEVPT2 method and its quasidegenerate version is presented. 
In Sec.~\ref{sec:qed}, the QED extension of both NEVPT2 and QD-NEVPT2 methods is discussed. 
In Sec.~\ref{sec:results}, the developed methods are tested on some particularly challenging multireference systems and compared against other ab initio methodologies.
We conclude the work with future perspectives presented in Sec.~\ref{sec:conclusions}.
\section{\label{sec:electronic} N-Electron Valence Perturbation Theory}
\subsection{\label{sec:electronic_NEVPT2} NEVPT2}
In NEVPT2, perturbation theory is performed on top of a CAS-type (CASCI, CASSCF, etc.) wavefunction which acts as zeroth-order reference:
\begin{equation}
   \ket{\Psi^{(0)}_{I}} = \sum_\Phi C_\Phi \ket{\Phi} \text{.}
\end{equation}
It is expressed as a linear combination of determinants $\ket{\Phi}$ with $C_\Phi$ being the expansion (CI) coefficients for the electronic state $I$.
Within the CASSCF approach, the orbital space is partitioned into three categories:
inactive orbitals that are always doubly occupied (denoted with indices $i, j, k, \dots$);
active orbitals characterized by non-integer occupations (indices $u, v, w, \dots$) and
virtual orbitals that are always unoccupied (indices $a, b, c, \dots$).
By distributing $n$ active electrons, within $m$ active orbitals, namely CAS($n,m$), the determinants are obtained and the CI coefficients are optimized until convergence.
Additionally, the orbital coefficients are also optimized.

The electronic structure problem is governed by the non-relativistic, Born--Oppenheimer electronic Hamiltonian, which in second-quantized form reads
\begin{equation}\label{eq:helec}
    H = \sum_{pq} h_{pq} E_{pq} + \frac{1}{2}\sum_{pqrs} g_{pqrs}\, e_{pqrs} ,
\end{equation}
where indices $p,q,r,s$ run over the full set of (inactive, active, and virtual) spatial orbitals, $h_{pq}$ are the one-electron integrals, and $g_{pqrs}$ are the two-electron repulsion integrals in chemists' notation. 
The operators $E_{pq}$ and $e_{pqrs}$ are the spin-traced one- and two-electron operators,
\begin{align}
    E_{pq} &= \sum_{\sigma} a^{\dagger}_{p\sigma} a_{q\sigma} , \\
    e_{pqrs} &= E_{pq}E_{rs} - \delta_{qr}E_{ps} = \sum_{\sigma\tau} a^{\dagger}_{p\sigma} a^{\dagger}_{r\tau} a_{s\tau} a_{q\sigma} ,
\end{align}
where $\sigma,\tau \in \{\alpha,\beta\}$ label the electron spin and $a^{\dagger}_{p\sigma}$ ($a_{p\sigma}$) creates (annihilates) an electron of spin $\sigma$ in spatial orbital $p$.

\begin{figure}[h!]
    \centering
    \includegraphics[width=\linewidth]{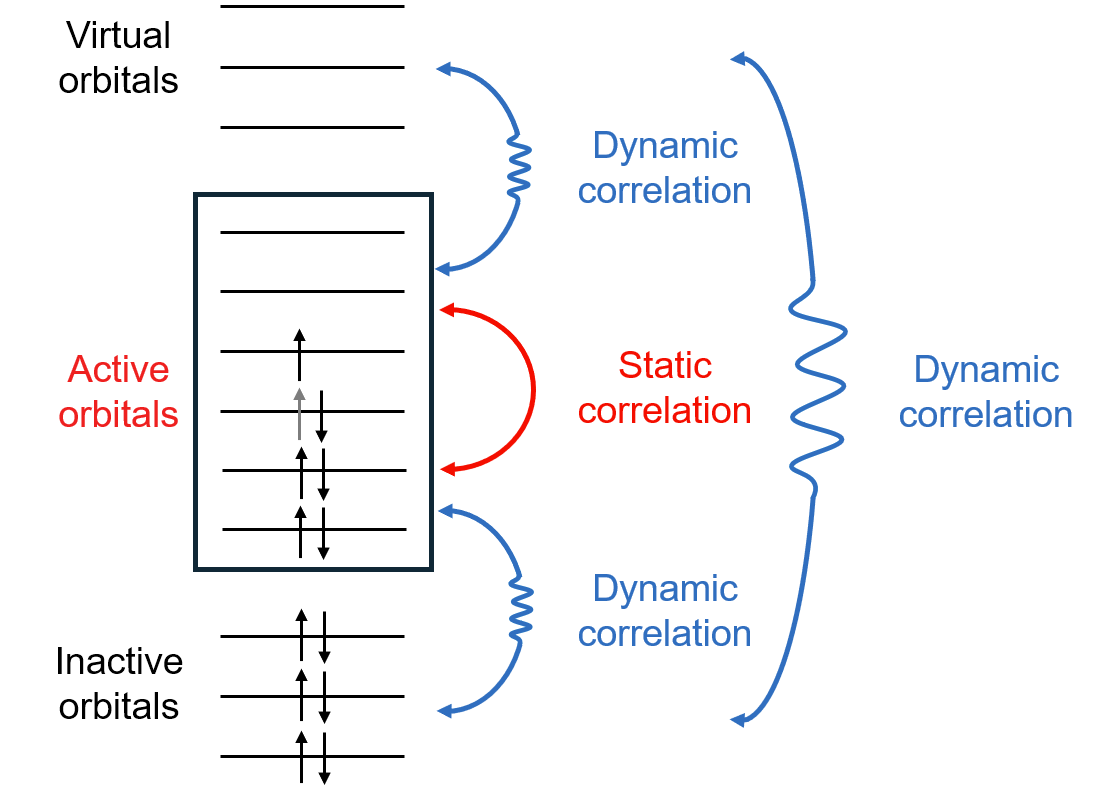}
    \caption{Graphical representation of the active space and of the excitations described by MRPT. If the active space is chosen properly, static correlation is predominantly captured by the active-active space interactions, while dynamic correlation is provided through interactions with the outer spaces.}
    \label{fig:Introduction_Fig2_PT_ActiveSpace}
\end{figure}

CASSCF recovers static correlation by expanding the wavefunction within a reference, or active, space. 
On top of this treatment, MRPT methods account for dynamic correlation by introducing excitations outside the active space, generally referred to as outer space (see Fig.~\ref{fig:Introduction_Fig2_PT_ActiveSpace}).
NEVPT2 is solved by a \textit{diagonalize-then-perturb} approach where the zeroth-order states are first obtained by diagonalizing the Hamiltonian within the active space and dynamic correlation is subsequently introduced perturbatively through excitations involving the outer space~\cite{andersson1990second, andersson1992second, battaglia2023multiconfigurational, chattopadhyay2016state}.

The Dyall Hamiltonian~\cite{dyall1995choice}, used as model Hamiltonian in NEVPT2 to build the zeroth-order Hamiltonian, is defined as:
\begin{equation}
    H^D = H_i + H_v + C ,
\end{equation}
where
\begin{equation}
    H_i = \sum_{i} \epsilon_{i} E_{ii} +
                \sum_{a} \epsilon_{a} E_{aa} ,
\end{equation}
is the one-electron term associated to the inactive and virtual spaces respectively, and $\epsilon_{i}$ and $\epsilon_{a}$ are the energies of the canonical orbitals.
\begin{equation}\label{hvdyall}
    H_v = \sum_{uv} h_{uv}^{eff} E_{uv} + \frac{1}{2}
                \sum_{uvwx} g_{uwvx} 
                (E_{uw} E_{vx} - \delta_{vw} E_{ux}) ,
\end{equation}
is the two-electron ``valence'' Hamiltonian acting within the active space, where
\begin{equation}\label{heff}
    h_{uv}^{eff} = h_{uv} + \sum_i (2g_{uvii} - g_{uiiv}) ,
\end{equation}
represents the effective one-electron matrix elements.
Finally
\begin{equation}
    C = 2\sum_i h_{ii} + \sum_{ij} (2g_{iijj} - g_{ijji}) - 2\sum_i \epsilon_i ,
\end{equation}
is a constant which ensures equivalence with the full Hamiltonian in the CAS space:
\begin{equation}
    P_{CAS} H P_{CAS} \ket{\Psi_{I} ^{(0)}} = H^D \ket{\Psi_{I} ^{(0)}} = E_{I} ^{(0)}\ket{\Psi_{I} ^{(0)}} .
    \label{eq:diag}
\end{equation}
 Here $P_{CAS}$ represents the projector onto the CAS space and $E_{I} ^{(0)}$ is the CASSCF (or CASCI) energy. 
By explicitly capturing two-body active-space interactions, this choice naturally removes the intruder-state problem.
This model Hamiltonian is then projected onto the first-order interacting space (FOIS)~\cite{meyer1977configurationFOIS}, spanned by contracted functions $\ket{K}$, generated by single and double excitations. The FOIS defines the space that provides first-order corrections to the reference wave function~\cite{mclean1973classification}. In this context, the contracted functions are linear combinations of configuration state functions (CSFs) which are orthogonal to $\ket{\Psi^{(0)}_{I}}$. 
These CSFs are grouped into subspaces denoted by $S^{(k)}_l$ where each subspace forms a complete active space itself, which ensures the invariance of the energy under rotations between active orbitals. 

The NEVPT2 method is classified into different variants based on the degree of contraction of the FOIS, which determines the number of perturber functions selected from each subspace~\cite{Angeli2001}.
The totally uncontracted variant retains all individual perturber functions, while the partially-contracted (PC) variant constructs perturber functions from the eigenstates of Dyall's valence Hamiltonian in the active space, retaining several contracted states per subspace. 
In contrast, the strongly-contracted (SC) variant retains a single contracted perturber function for each subspace.

In the SC variant, the one adopted in this work, each contracted function is obtained by applying a perturber operator $V_l^{(k)}$, which acts as a linear combination of double excitation operator, to the zeroth-order wavefunction: 
\begin{equation}
    \ket{K} = {P}_{S_l^{(k)}} {H} \ket{\Psi_{I} ^{(0)}} 
    = V_l^{(k)} \ket{\Psi_{I} ^{(0)}} ,
\end{equation}
where $k$ denotes the number of electrons promoted to or removed from the active space, $l$ specifies the fixed occupation pattern of the inactive orbitals, and $P_{S_l^{(k)}}$ is the projector onto subspace $S_l^{(k)}$~\cite{Angeli2002}. 

In SC-NEVPT2, eight distinct classes of perturbers are defined~\cite{Angeli2001, angeli2001n, Angeli2002, angeli2007new}:
\begin{subequations} \label{perturbatorsnev}
        \begin{align}
        & {V}^{(0)}_{ij,ab} = \eta_{ab}\eta_{ij} [(ai | bj)E_{ai}E_{bj} + (bi | aj)E_{bi}E_{aj}], \\
        & {V}^{(+1)}_{ij,a} = \eta_{ij}\sum_{u} [( ai | uj) E_{ai}E_{uj} + (aj | ui)E_{aj}E_{ui}] , \\
        & {V}^{(-1)}_{ab, i} = \eta_{ab}\sum_{u} [(ai | bu)E_{ai}E_{bu} + (bi | au) E_{bi}E_{au}] ,\\
        & {V}^{(+2)}_{ij} = \eta_{ij}\sum_{uv}( ui | vj)E_{ui}E_{vj} , \\
        & {V}^{(-2)}_{ab} = \eta_{ab}\sum_{uv}( au | bv)E_{au}E_{bv} , \\
        & {V}^{(0)'}_{i,a} = \sum_{uv}[(ai|uv)E_{ai}E_{uv} + (ui|av)E_{ui}E_{av} ] + h_{ai}^{eff}E_{ai} , \\
        & {V}^{(+1)'}_{i} = \sum_{uvw}(ui | wv) E_{ui}E_{wv} + \sum_{u} h_{ui}^{eff}E_{ui} ,\\
        & {V}^{(-1)'}_{a} = \sum_{uvw}( au| wv)E_{au}E_{wv} +\sum_{u} h_{au}^{eff'}E_{au} ,
    \end{align}
\end{subequations}
with $\eta_{pq} = 1 - \frac{1}{2} \delta_{pq}$ and
\begin{equation}
    h_{pq}^{eff'} = h_{pq}^{eff} - \sum_{v} g_{pvvq} \text{.}
\end{equation}
It is important to point out that the perturber functions $\ket{K}$ are not normalized to unity but are orthogonal~\cite{Angeli2001, angeli2001n}.


The zeroth order energies of the perturbers in the classes introduced previously can be associated to well-defined physical processes occurring within the active space, a correspondence that is ultimately responsible for the absence of intruder states in NEVPT2~\cite{angeli2007new}. 

%
%
From a computational point of view, the evaluation of these energies requires density matrices of increasing rank, all restricted to active orbital indices. 
Most classes only involve up to the three-body reduced density matrix (RDM3), while the $V^{(-1)'}_{a}$ and $V^{(+1)'}_{i}$ classes are the most demanding, requiring auxiliary quantities built from the four-body reduced density matrix (RDM4), which is the main computational bottleneck of NEVPT2 calculations~\cite{angeli2007new}. 
Several strategies have been proposed to avoid its explicit evaluation, ranging from cumulant approximations~\cite{nakatsuji1976equation} and matrix-product-state compression~\cite{sharma2017combining} to imaginary-time propagation~\cite{sokolov2016time, sokolov2017time} and Monte Carlo sampling~\cite{mahajan2019multireference}. 
Among these, factorizing the RDM4 contributions in terms of lower-rank intermediates, obtained by pre-contracting the active-space two-electron integrals with the reference wavefunction, is particularly appealing, as it lowers the computational scaling without introducing any additional approximation~\cite{sokolov2024multireference}.
This is the strategy adopted in this work.
\subsection{\label{sec:electronic_QD} Quasidegenerate NEVPT2}
While state-specific NEVPT2 can be applied to individual ground or excited states, it does not account for the coupling between two or more states when they are close in energy.
Neglecting such interaction can lead to an incorrect description of potential energy surfaces near conical intersections, avoided crossings, or in regions with a high density of states.
In the quasidegenerate (QD) formulation of NEVPT2~\cite{angeli2004quasidegenerate}, this interaction between close states  is recovered by diagonalizing an effective Hamiltonian built within the space $\mathcal{M}$ of quasidegenerate CASSCF states, following the \textit{"perturb-then-diagonalize"} scheme.
In this approach the perturbation is applied before final diagonalization and the effective Hamiltonian is defined as:
\begin{equation}\label{eq:heff_bloch}
H_{\rm eff} \ket{\tilde{\Psi}_{I}} = E_{I} \ket{\tilde{\Psi}_{I}} ,  H_{\rm eff} = PH\Omega ,
\end{equation}
with $P = \sum_{I \in \mathcal{M}} \lvert \Psi_I^{(0)} \rangle \langle \Psi_I^{(0)} \rvert$, and $\Omega$ the wave operator satisfying equation~\cite{bloch1958theorie}:
\begin{equation}
\Omega P H \Omega - H \Omega = 0 .
\end{equation}
Following Lindgren's formulation~\cite{lindgren1974rayleigh}, the effective Hamiltonian and the wave operator are expanded order by order in the perturbation 
\begin{equation}
\Omega = P + \Omega^{(1)} + \Omega^{(2)} + \dots ,
\end{equation}
\begin{equation}
H_{\rm eff} = H_{\rm eff}^{(0)} + H_{\rm eff}^{(1)} + H_{\rm eff}^{(2)} + \dots ,
\end{equation}
yielding the working expressions up to second order:
\begin{align}
H_{\rm eff}^{(0)} &= P H_0 P , \\
H_{\rm eff}^{(1)} &= P V P = 0 , \\
H_{\rm eff}^{(2)} &= P V \Omega^{(1)} .
\end{align}
To obtain a practical evaluation of $H_{\rm eff}^{(2)}$, the multipartitioning technique of Zaitsevskii and Malrieu~\cite{zaitsevskii1995multi} $H = H_0(I)+V(I)$ is adopted, introducing state-dependent partitionings of the Hamiltonian. As consequence, the matrix elements of $H_{\rm eff}$ between two states $I, J \in \mathcal{M}$ up to second order read as
\begin{align}
H^{(0)}_{{\rm eff},IJ} &= E_I^{(0)} \delta_{IJ} , \\
H^{(1)}_{{\rm eff},IJ} &= \langle \Psi_J^{(0)} \rvert V \lvert \Psi_I^{(0)} \rangle = 0 , \\
H^{(2)}_{{\rm eff},IJ} &= \langle \Psi_J^{(0)} \rvert V \lvert \Psi_I^{(1)} \rangle ,
\end{align}
where $\ket{\Psi_I^{(1)}} = \Omega^{(1)} \ket{\Psi_I^{(0)}}$ is the first-order correction to the zeroth-order state, $I$.

This effective Hamiltonian is, in general, non-Hermitian~\cite{angeli2004quasidegenerate}: the perturber energies entering $H^{(2)}_{\rm eff}$ depend on the ket state $I$ but not on the bra state $J$, so $H^{(2)}_{{\rm eff},IJ}$ is not symmetric under $I \leftrightarrow J$, and the eigenvalues of $H_{\rm eff}$ are therefore not guaranteed to be real.

An Hermitian formulation can be obtained a posteriori via a des Cloizeaux transformation~\cite{des1960extension}, $H_{\rm eff}' = S^{-1/2} H_{\rm eff} S^{1/2}$ with $S_{IJ} = \langle \tilde{\Psi}_I \vert \tilde{\Psi}_J \rangle$, or directly through canonical Van Vleck perturbation theory~\cite{shavitt1980quasidegenerate, kirtman1981simultaneous}, where the second-order term is symmetrized as
\begin{equation}
H^{(2)}_{{\rm eff},IJ} = \frac{1}{2} \left( \langle \Psi_J^{(0)} \rvert V \lvert \Psi_I^{(1)} \rangle + \langle \Psi_J^{(1)} \rvert V \lvert \Psi_I^{(0)} \rangle \right) .
\end{equation}
The two formulations typically yield very similar results, but the Hermitian one guarantees real energies and, thanks to the symmetric eigenvalue problem, simplifies the evaluation of molecular properties~\cite{sokolov2024multireference}.
\section{\label{sec:qed} QED-NEVPT2}
The QED-CASSCF wavefunction for a general polaritonic state $I$ can be written as
\begin{equation}
    \ket{\Psi _{I}} = \sum_\Phi \sum_m C_{\Phi,m} \ket{\Phi} \otimes \ket{m} ,
\end{equation}
where $\ket{\Phi}$ is an electronic determinant and $\ket{m}$ is the $m^{th}$ photonic state of the cavity mode.

The coupled electron-photon system, in ab-initio polaritonic quantum chemistry methods, can be described by the nonrelativistic single-mode Pauli-Fierz Hamiltonian in length gauge and dipole approximation:
\begin{equation}\label{eq: Starting_PF}
    \begin{split}
    H_{PF} &= H_e +
    \omega b^{\dag}b + 
    \frac{1}{2}\Big[\boldsymbol{\lambda} \cdot (\boldsymbol{d} - \langle \boldsymbol{d} \rangle )
    \Big]^2 \\
    &- 
    \sqrt{\frac{\omega}{2}}\Big[\boldsymbol{\lambda} \cdot \left(\boldsymbol{d} - \langle \boldsymbol{d} \rangle \right)\Big](b^{\dag} + b) \text{.}
    \end{split}
\end{equation}
In the equation above, $H_e$ denotes the standard electronic Hamiltonian under the Born-Oppenheimer approximation, while $\omega b^\dagger b$ corresponds to the purely photonic term. The subsequent terms account for cavity QED effects: the dipole self-energy (DSE) contribution, $\frac{1}{2}(\boldsymbol{\lambda} \cdot (\boldsymbol{d} - \langle \boldsymbol{d} \rangle))^2$, ensures that the Hamiltonian remains bounded from below~\cite{rokaj2018light}, while the final term describes the electronic-photonic bilinear coupling. Here, $\boldsymbol{d}$ denotes the dipole moment operator, with $\langle \boldsymbol{d} \rangle$ representing its expectation value, and $\boldsymbol{\lambda}$ incorporates both the polarization vector and the cavity quantization volume.
For simplicity, our derivation focuses on a single cavity mode, though its extension to multimode is straightforward. To ensure origin invariance, Eq.~\ref{eq: Starting_PF} has already been expressed in the coherent-state basis~\cite{klauder1985coherent}. Adopting the notation used in~\cite{lexander2024analytical} , we can conveniently set $d = \boldsymbol{\lambda}\cdot\boldsymbol{d}$ (and analogously $\langle d\rangle = \boldsymbol{\lambda}\cdot\langle\boldsymbol{d}\rangle$), and rewrite the Pauli–Fierz Hamiltonian as:
\begin{equation}\label{eq:PF_Hamiltonian}
    \begin{split}
    {H}_{PF} &= \sum_{pq} h_{pq} {E}_{pq}
    + \frac{1}{2} \sum_{pqrs} g_{pqrs} {e}_{pqrs} \\
    &+ \omega {b}^{\dagger} {b}
    + \sqrt{\frac{\omega}{2}} \sum_{pq} d_{pq} {E}_{pq} ({b}^{\dagger} + {b})
    - \sqrt{\frac{\omega}{2}} \langle {d} \rangle ({b}^{\dagger} + {b}) \text{.}
    \end{split}
\end{equation}
For compactness, we have absorbed the dipole self-energy (DSE), into the one- and two-electron integrals:
\begin{equation}
    \begin{split}
        & h_{pq} = h^e _{pq} + \frac{1}{2} q_{pq} - d_{pq}\langle d \rangle + \frac{\delta_{pq}}{2N_e} \langle d \rangle ^2 , \\
        & g_{pqrs} = g^e _{pqrs} + d_{pq}d_{rs} ,
    \end{split}
\end{equation}
where $d_{pq}$ and $q_{pq}$ are the electric dipole and quadrupole integrals, respectively, and $N_e$ is the number of electrons in the molecule.

Extending NEVPT2 theory to this framework, requires a new zeroth-order Hamiltonian that would accounts also for the photonic and light-matter interaction terms of the total Hamiltonian $H_{PF}$. 
We therefore introduce the QED-Dyall Hamiltonian:
\begin{equation}
    H^D_{QED} = H_i + H_v + H_{ph} + C ,
\end{equation}
where
\begin{equation}
    \begin{split}
    H_i &= \sum_{i} \epsilon_{i} E_{ii} +
            \sum_{a} \epsilon_{a} E_{aa} \\
            &+ \sqrt{\frac{\omega}{2}} \left(\sum_{i}  d_{ii}E_{ii} + \sum_a d_{aa} E_{aa} \right) (b^{\dagger} + b) ,
    \end{split}
\end{equation}
and
\begin{equation}
    \begin{split}
    H_v &= \sum_{uv} h_{uv}^{eff} E_{uv} +
        \frac{1}{2} \sum_{uvxy} g_{uvxy} e_{uvxy} \\&+ \sqrt{\frac{\omega}{2}} \sum_{uv} d_{uv}E_{uv} (b^{\dagger} + b) ,
    \end{split}
\end{equation}
are the new one-electron inactive term and two-electron valence Hamiltonian, respectively, extended to include the bilinear coupling. 
The orbital energies $\epsilon_i$ and $\epsilon_a$ are the eigenvalues of the inactive and virtual blocks of the generalized Fock matrix of the QED-CASSCF state under study (state-specific canonicalization) without the bilinear coupling term.
\begin{equation}
    H_{ph} = \omega b^{\dagger}b ,
\end{equation}
is the purely photonic term, and $C$ maintains the same form as electronic NEVPT2 theory, with the DSE-dressed one- and two-electron integrals. 
%
Unlike the purely electronic Dyall Hamiltonian, $H_i$ and $H_v$ now carry an explicit dependence on the photon creation and annihilation operators through the bilinear coupling term, while $H_{ph}$ is a genuinely new, stand-alone contribution accounting for the cavity mode. 
By construction, $H^D_{QED}$ reproduces the exact Pauli-Fierz Hamiltonian energy within the QED-CAS space,
\begin{equation}
    P_{QCAS} H_{PF} P_{QCAS} \ket{\Psi_{I}^{(0)}} = 
    H^D_{QED} \ket{\Psi_{I}^{(0)}} = 
    E_{I}^{(0)} \ket{\Psi_{I}^{(0)}} ,
\end{equation}
where $P_{QCAS}$ is the projector onto the QED-CAS space and $E_{I}^{(0)}$ is the QED-CAS energy of state $I$. 
Projecting $H^D_{QED}$ onto the FOIS defines the QED zeroth-order Hamiltonian. 

\begin{figure*}[t]
    \centering
    \includegraphics[width=1.0\linewidth]{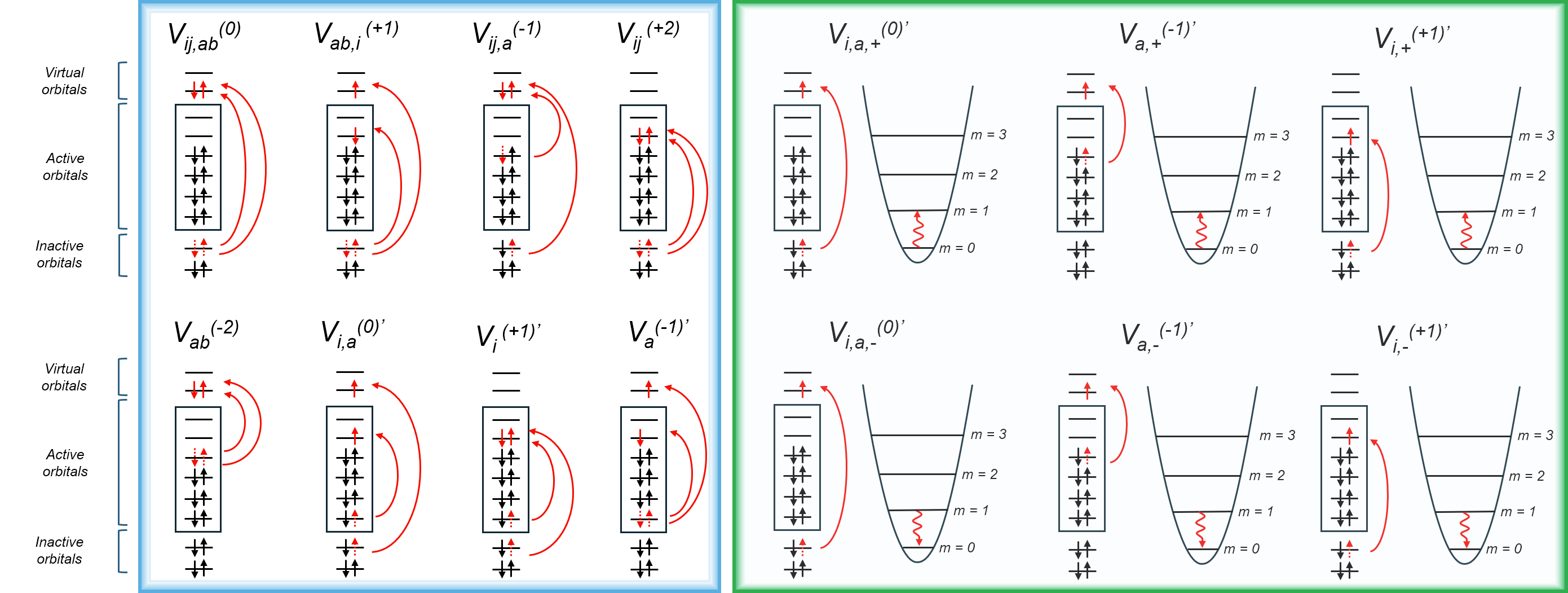}
    \caption{Graphical representation of the purely electronic perturbers $\hat{V}^{(k)}_l$ (blue box) and of the mixed electron-photon perturbers $\hat{V}^{(k)}_{l,+}$ and $\hat{V}^{(k)}_{l,-}$ (green box).}
    \label{fig:Introduction_Fig3_Perturbers}
\end{figure*}

It is crucial to highlight that the explicit inclusion of the field, enlarges the FOIS. Now, it can be conceptually separated into three orthogonal subspaces according to the change in photon number (Fig.~\ref{fig:Introduction_Fig3_Perturbers}):
\begin{equation}
    \begin{split}
    &\ket{K} = V^{(k)}_l \ket{\Psi_{I}^{(0)}} ,\\
    &\ket{K_+} = V^{(k)}_{l,+} \ket{\Psi_{I}^{(0)}} ,\\
    &\ket{K_-} = V^{(k)}_{l,-} \ket{\Psi_{I}^{(0)}} ,
    \end{split}
\end{equation}

corresponding to no photon exchange, photon creation, and photon annihilation, respectively. 
The $\ket{K}$ functions coincide with the eight standard NEVPT2 classes~\cite{angeli2001n}, while $\ket{K_+}$ and $\ket{K_-}$ arise from the bilinear coupling term and involve a single electronic excitation coupled to a photon creation or annihilation. Because $H_{PF}$ contains photon operators exclusively through the number operator $b^\dagger b$ and the bilinear term $b^\dagger + b$, double-photon creation or annihilation processes are absent. Consequently, no perturbers can change the photon number by two, restricting mixed perturbers to those coupling a single electronic excitation with a single photon. Each of the $\ket{K_+}$ and $\ket{K_-}$ further split into three classes, depending on the orbital spaces involved in the electronic excitation; the two sets of classes are structurally identical and differ only in the photonic part.
Starting from $\ket{K_+}$, the three classes correspond to excitations purely within the inactive/virtual space, from an active orbital into a virtual orbital, and from an inactive orbital into an active orbital, respectively:
\begin{subequations}
    \begin{align}
        & V_{i,a,+}^{(0)'} = \sqrt{\frac{\omega}{2}}  d_{ai} E_{ai} b^{\dagger} , \\
        & V_{i,+}^{(+1)'} = \sqrt{\frac{\omega}{2}} \sum_{u} d_{ui} E_{ui} b^{\dagger} ,  \\
        & V_{a,+}^{(-1)'} =  \sqrt{\frac{\omega}{2}} \sum_{u} d_{au} E_{au} b^{\dagger} \text{.}
    \end{align}
\end{subequations}
The corresponding operators for $\ket{K_-}$ are obtained by replacing $b^{\dagger}$ with $b$, while keeping the same electronic contributions:
\begin{subequations}
    \begin{align}
        & V_{i,a,-}^{(0)'} = \sqrt{\frac{\omega}{2}} d_{ai} E_{ai} b , \\
        & V_{i,-}^{(+1)'} = \sqrt{\frac{\omega}{2}}\sum_{u} d_{ui} E_{ui} b , \\
        & V_{a,-}^{(-1)'} = \sqrt{\frac{\omega}{2}} \sum_{u} d_{au} E_{au} b.
    \end{align}
\end{subequations}
As for the electronic case, these contracted functions are not normalized but remain mutually orthogonal, so that they do not interact with the reference wavefunction via the total Hamiltonian.

To obtain the perturbative corrections to the energy we follow the spin-free formalism of Angeli et al.~\cite{Angeli2002}, which avoids the computation of density matrices of orders higher than the fourth. Within this approach, the energy of each perturber, $E_l^{(k)}$, is not obtained from an explicit evaluation of the matrix elements $\matrixel{K}{H^D_{QED}}{K}$, but from the expectation value of the commutator between the Dyall Hamiltonian and the perturber operator. This is the same strategy underlying the extended Koopmans' theorem~\cite{smith1975extension}, in which ionization potentials and electron affinities are expressed as expectation values of commutators of the Hamiltonian with a creation or annihilation operator, rather than as energy differences between the neutral and ionized states.

Within the strongly contracted QED-NEVPT2 formalism, the second-order energy correction for the state ${I}$ is obtained, as in the electronic case, by summing the contributions of all subspaces,
\begin{equation}
    \label{eq:starting_eq}
    E_{I}^{(2)} = \sum_{k} \sum_{l} \frac{N_l^{(k)}}{E_{I}^{(0)} - E_l^{(k)}} ,
\end{equation}
where $N_l^{(k)} = \langle \Psi_{I}^{(0)} \rvert ({V}_l^{(k)})^{\dagger} {V}_l^{(k)} \lvert \Psi_{I}^{(0)} \rangle$ is the norm of the corresponding perturber, and
\begin{equation}
    E_l^{(k)} = \frac{1}{N_l^{(k)}} \langle \Psi_{I}^{(0)} \rvert ({V}_l^{(k)})^{\dagger} {H}^D_{QED} {V}_l^{(k)} \lvert \Psi_{I}^{(0)} \rangle .
\end{equation}
is the corresponding zeroth order energy.

Using the commutator identity $A B = [A, B] + B A$ and Eq.~\ref{eq:diag}, the energy correction for each subspace can be written as
\begin{equation}
    E_l^{(k)} - E_{I}^{(0)} = \frac{1}{N_l^{(k)}} \langle \Psi_{I}^{(0)} \rvert ({V}_l^{(k)})^{\dagger} [{H}^D_{QED}, {V}_l^{(k)}] \lvert \Psi_{I}^{(0)} \rangle .
\end{equation}

Until this point, we referred to $E_l^{(k)}, N_l^{(k)} \text{and}\; {V}_l^{(k)}$ as generic zeroth order energies, normalization factors and perturbers' operators, respectively, belonging to all the subspaces.
From now on, these terms will refer to the $\ket{K}$ class.
Explicit references to the energies, normalizations factors and perturbers' operators of the newly introduced subspaces $\ket{K_{\pm}}$ will carry the additional $\pm$ notation.

For the purely electronic class $\ket{K}$, $[ H_{ph},  V_l^{(k)}] = 0$, as these operators do not act on the photonic degrees of freedom, and the excitation energy reduces to the same two-term form as in standard NEVPT2,
\begin{equation}
    E_l^{(k)} - E_{I}^{(0)} = \Delta\epsilon_l^{(k)} + \xi_l^{(k)} ,
\end{equation}
with
\begin{equation}
    \Delta\epsilon_l^{(k)} = 
    \frac{1}{N_l^{(k)}} \langle \Psi_{I}^{(0)} | ({V}_l^{(k)})^{\dagger} [{H}_i, {V}_l^{(k)}] | \Psi_{I}^{(0)} \rangle  ,
\end{equation}
\begin{equation}
    \xi_l^{(k)} = 
    \frac{1}{N_l^{(k)}} \langle \Psi_{I}^{(0)} | ({V}_l^{(k)})^{\dagger} [{H}_v, {V}_l^{(k)}] | \Psi_{I}^{(0)} \rangle ,
\end{equation}
For the mixed light-matter classes $\ket{K_+}$ and $\ket{K_-}$, the perturbation operators $V_{l,\pm}^{(k)}$ carry a partial bosonic character; therefore, the photonic commutator gives and additional term:
\begin{equation}\label{eq:starting_plus_minus}
    \frac{1}{N_{l,\pm}^{(k)}} \langle \Psi_{I}^{(0)} \rvert ({V}_{l,\pm}^{(k)})^{\dagger} [{H}_{ph}, {V}_{l,\pm}^{(k)}] \lvert \Psi_{I}^{(0)} \rangle ,
\end{equation}
%

As a representative example, we show the results obtained for the subspace $S_{a,+}^{(-1)'}$, explicit formulas for the other perturbers are reported in Appendix~\ref{AppendixA},~\ref{AppendixB} and~\ref{AppendixC}.

The final contribution to the second-order energy for the state I of this subspace is given by:
\begin{equation}
    E^{(2)}_{I}(S_{a,+}^{(-1)'})=- \frac{N_{a,+}^{(-1)'}}{\Delta\epsilon_{a,+}^{(-1)'} + \xi_{a,+}^{(-1)'} + \omega} \quad \text{.} 
\end{equation} 
To evaluate this expression, the constituent terms are defined as follows. First, the numerator is given by:
\begin{equation}
    \begin{split}
    N_{a,+}^{(-1)'} &= \frac{\omega}{2} \sum_{u'u} d_{au'} d_{au} \sum_{\Phi,\Theta} \sum_{m} C_{\Phi,m}\\
    &\times  \matrixel{\Phi}{E_{u'u}}{\Theta} C_{\Theta,m} (m+1) .
    \end{split}
\end{equation}
The inactive term $\Delta\epsilon_{a,+}^{(-1')}$ is defined as:
\begin{equation}
    \begin{split}
    &\Delta\epsilon_{a,+}^{(-1)'} = \frac{1}{N_{a,+}^{(-1)'}} \langle \Psi^{(0)}_{I} | ({V}_{a,+}^{(-1)'})^{\dagger} [{H}_i, {V}_{a,+}^{(-1)'}] | \Psi^{(0)}_{I} \rangle\\
    & = \epsilon_a  + \left( \frac{\omega}{2} \right)^{\frac{3}{2}} \sum_{u'u} d_{au'} d_{au} \Bigg\{  2\sum_i d_{ii} \\
    & \times \sum_{\Phi,\Theta} \sum_{m} C_{\Phi,m-1} \matrixel{\Phi}{E_{u'u}}{\Theta} C_{\Theta,m} \sqrt{m}\\
    &+  d_{aa} \Big[ \sum_{\Phi,\Theta} \sum_{m} C_{\Phi,m+1} \matrixel{\Phi}{E_{u'u}}{\Theta} C_{\Theta,m} \sqrt{m+1} (m + 2) \\
        &+\sum_{\Phi,\Theta} \sum_{m} C_{\Phi,m-1} \matrixel{\Phi}{E_{u'u}}{\Theta} C_{\Theta,m} \sqrt{m} (m + 1)\Big] \Bigg\} / N_{a,+}^{(-1)'} \text{.}
    \end{split}
\end{equation}
The contribution from the valence term of the QED-Dyall Hamiltonian appears as:
\begin{equation}
    \begin{split}
    &\xi_{a,+}^{(-1)'} = \frac{1}{N_{a,+}^{(-1)'}} \langle \Psi^{(0)}_{I} \rvert (V_{a,+}^{(-1)'})^{\dagger} [H_{v}, V_{a,+}^{(-1)'}] \lvert \Psi^{(0)}_{I} \rangle \\
    &= \frac{\omega}{2 N_{a,+}^{(-1)'}} \sum_{\Phi,\Theta} \sum_{m} \sum_{u'u} d_{au'} d_{au} \\
    &\times \Bigg\{
    - \sum_{w} h^{eff}_{uw} C_{\Phi,m} \matrixel{\Phi}{E_{u'w}}{\Theta} C_{\Theta,m} (m + 1) \\
    &- \sum_{wxy} g_{uwxy} C_{\Phi,m} \matrixel{\Phi}{E_{u'w}E_{xy}}{\Theta} C_{\Theta,m} (m + 1)\\
    &- \sqrt{\frac{\omega}{2}} \sum_w d_{uw} \Big[ C_{\Phi,m+1} \matrixel{\Phi}{E_{u'w}}{\Theta} C_{\Theta,m} \sqrt{m+1} (m + 2)  \\
    &+ C_{\Phi,m-1} \matrixel{\Phi}{E_{u'w}}{\Theta} C_{\Theta,m} \sqrt{m} (m + 1) \Big]\\
    &+ \sqrt{\frac{\omega}{2}} \sum_{vw} d_{vw} C_{\Phi,m-1} \matrixel{\Phi}{E_{u'u}E_{vw}}{\Theta} C_{\Theta,m} \sqrt{m}
    \Bigg\} \text{,}
    \end{split}
\end{equation}
while the photonic contribution is given by:
\begin{equation}
    \begin{split}
    & \frac{1}{N_{a,+}^{(-1)'}} \langle \Psi^{(0)}_{I} \rvert ({V}_{a,+}^{(-1)'})^{\dagger} [{H}_{ph}, {V}_{a,+}^{(-1)'}] \lvert \Psi^{(0)}_{I} \rangle = \omega \text{.}
    \end{split}
\end{equation}
Several perturber classes of $\ket{K}$, namely those with $k=\pm1$ and $k=\pm2$, involve auxiliary matrices defined in the active space. These matrices are the spin-free analogues of those arising in the extended Koopmans' theorem~\cite{smith1975extension,Morrell1975} (Eqs.~\eqref{eq:koopmans-1}, \eqref{eq:koopmans+1}, \eqref{eq:koopmans-2} and \eqref{eq:koopmans+2}), and their explicit expressions are reported in Appendix~\ref{AppendixA}.

In standard NEVPT2~\cite{Angeli2002}, these matrices describe well-defined processes within the active space such ionization or electron attachment ($k=\pm1$) (or the corresponding double processes for $k=\pm2$), and their diagonalization yields approximate ionization potentials and electron affinities. 

Unlike the purely electronic case, however, the notion of ionization in a polaritonic systems is inherently ambiguous~\cite{deprince2021cavity, riso2022characteristic}, and therefore, none of these Koopmans-like matrices can be unambiguously associated with a well-defined (photo)ionization process, and we leave a detailed interpretation of these terms to future work. 

The evaluation of the QED-NEVPT2 energy contributions requires reduced density matrices of the same order as those of electronic NEVPT2, computed on the QED wavefunction.

Within $\ket{K}$, the $0'$ and $\pm1'$ classes remain the most computationally demanding, exactly as in standard NEVPT2; the $\pm1'$ classes additionally involve, alongside the RDM4, a three-body bilinear density matrix, the highest-rank bilinear matrix appearing in this work (Eqs.~\ref{eq:4rdm-1} and~\ref{eq:4rdm-2}).
The new perturbers $\ket{K_+}$ and $\ket{K_-}$, in contrast, contribute only marginally to the overall computational cost.

Although $\Psi^{(0)}_{I}$ now includes the photonic degrees of freedom, the RDM4 entering the $\pm1'$ classes of $\ket{K}$ is built from purely electronic excitation operators acting on active orbital indices; the photon number only enters implicitly, through the coefficients of $\Psi^{(0)}_{I}$, and does not introduce any additional index. 
The algebraic structure of the RDM4 is therefore unchanged with respect to the electronic case, and all the approximations discussed above, which act precisely on this structure, remain directly applicable to QED-NEVPT2.
\subsection{\label{sec:qd_qed_NEVPT2} QED-QD-NEVPT2}
The presence of the cavity can bring electronic states closer in energy, for instance, in the case where the Rabi splitting is formed, favoring the appearance of quasidegeneracies that would not occur outside the cavity environment. 
In this sense, the quasidegenerate treatment is not merely a refinement but becomes an essential tool for obtaining a correct description of polaritonic states.

The extension to the quasidegenerate case, QED-QD-NEVPT2, follows directly from the formalism outlined in Sec.~\ref{sec:electronic_QD}.
Since the effective Hamiltonian of Eq.~\eqref{eq:heff_bloch} depends only on the model space and on the perturbers, the QD treatment is formally identical to the electronic case, with $\ket{K}$, $\ket{K_+}$, and $\ket{K_-}$ now generated with the application of the appropriate linear combinations of the excitation operators of the QED Hamiltonian to each state $I$ of the model space.
The resulting effective Hamiltonian retains the same non-Hermitian structure discussed above, and can analogously be made Hermitian via a des Cloizeaux transformation or through canonical Van Vleck perturbation theory.
\section{\label{sec:results} Results}
In this section, QED-NEVPT2 and its quasidegenerate counterpart (QED-QD-NEVPT2) are applied to evaluate field-induced effects in systems characterized by non-negligible static and dynamic electron correlation. 
All computations were carried out using a development version of the $e^T$ program suite~\cite{folkestad20201,folkestad20252}. 
A single cavity mode is used throughout, with the light--matter coupling strength set to $\lambda = 0.05$~a.u. and the field polarization along the $(1,1,1)/\sqrt{3}$ direction.
Reference molecular geometries, together with additional comparisons and further analysis of the perturber contributions are reported in the Supporting Information (SI).
\subsection{\label{sec:results_ammonia} Ammonia}
\begin{figure}[h!]
    \centering
    \includegraphics[width=0.5\linewidth]{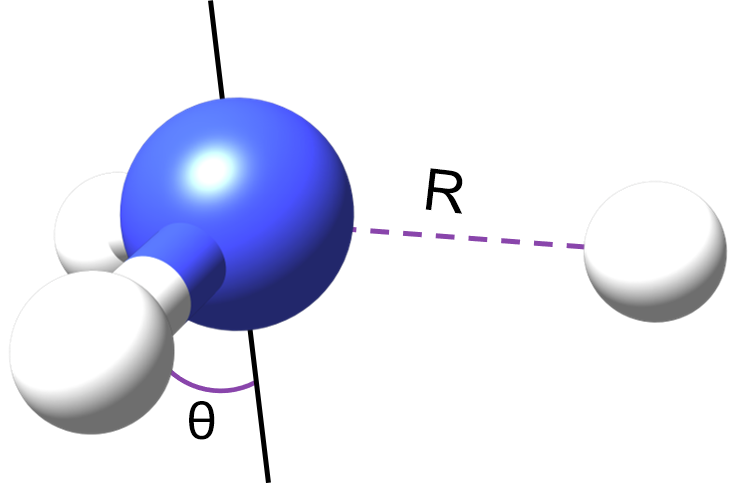}
    \caption{Representation on the $R$ and $\theta$ parameters of ammonia.}
    \label{fig:Ammonia_Fig1_MoleculeGeometry}
\end{figure}

The first analysis performed in this paper involves the excited-state potential energy surfaces of ammonia. 
In this system, an $S_0/S_1$ conical intersection is reached if the N--H bond is stretched in the planar geometry.~\cite{rossi2026convex}
The NH$_3$ potential energy surfaces has been inspected by varying the distance $R$ of one N--H bond taking the other two fixed at the equilibrium distance (1.04~\AA) and by changing the $\theta$ angle defined in Fig.~\ref{fig:Ammonia_Fig1_MoleculeGeometry}.
\noindent All calculations were performed with the 3-21G basis set. 
Using this basis set a direct comparison with QED-FCI reference energies across the entire potential energy surface was possible.
QED-QD-NEVPT2 calculations have been performed on top of QED-CASSCF with an active space of 8 electrons in 7 orbitals, CAS(8,7).
As an additional point of comparison, we also computed QED-CCSD energies, which is generally the reference approach for correlated molecular systems. 
However, it is know to fail when, as in the conical intersection case, more degenerate configurations need to be simultaneously taken into account.

\begin{figure}[h!]
    \centering
    \includegraphics[width=1\linewidth]{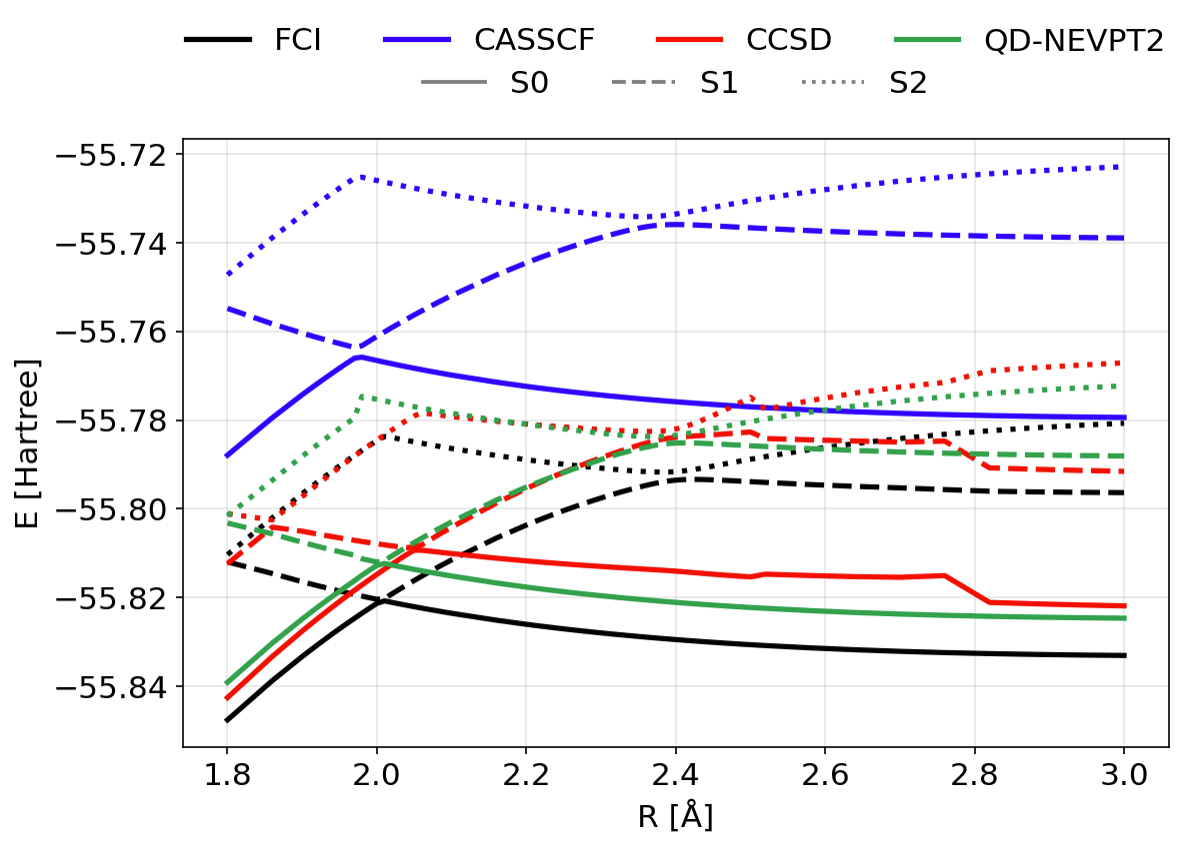}
    \caption{Polaritonic potential energy curves for the three lowest singlet states of ammonia at the $\theta=90^{\circ}$ geometry computed at the QED-FCI (black), QED-CASSCF(8,7) (blue), QED-CCSD (red), and QED-QD-NEVPT2 (green)  levels of theory.}
    \label{fig:Ammonia_Fig2_PES_QED_2D_90}
\end{figure}

For each method, the cavity frequency $\omega$ was set equal to the purely electronic $S_0 \rightarrow S_1$ excitation energy computed at $R = 2.4$~\AA{}.
Each QED calculation therefore uses its own method-specific frequency, so that the cavity is resonant with the electronic transition at that geometry for every level of theory.
We first focus on the $\theta=90^{\circ}$ case, which corresponds to the planar geometry at which the conical intersection is present. 
In Fig.~\ref{fig:Ammonia_Fig2_PES_QED_2D_90}, the three lowest singlet states are plotted for QED-QD-NEVPT2, QED-CASSCF, QED-CCSD, and QED-FCI.
In this analysis, QED-NEVPT2 was not included since the method is known to fail near degeneracies.
In fact, the purely electronic NEVPT2 calculation for the lowest two electronic states already presents an unphysical double crossing located in the conical intersection region at $\sim$ 2 \AA{}. 
This behavior is well-known and documented~\cite{spiegelmann1984use1, spiegelmann1984use}, and the conical intersection is restored by relying on the QD-NEVPT2 method (see Fig.~S1).
%
%
\begin{figure*}[ht!]
    \centering
    \includegraphics[width=1\linewidth]{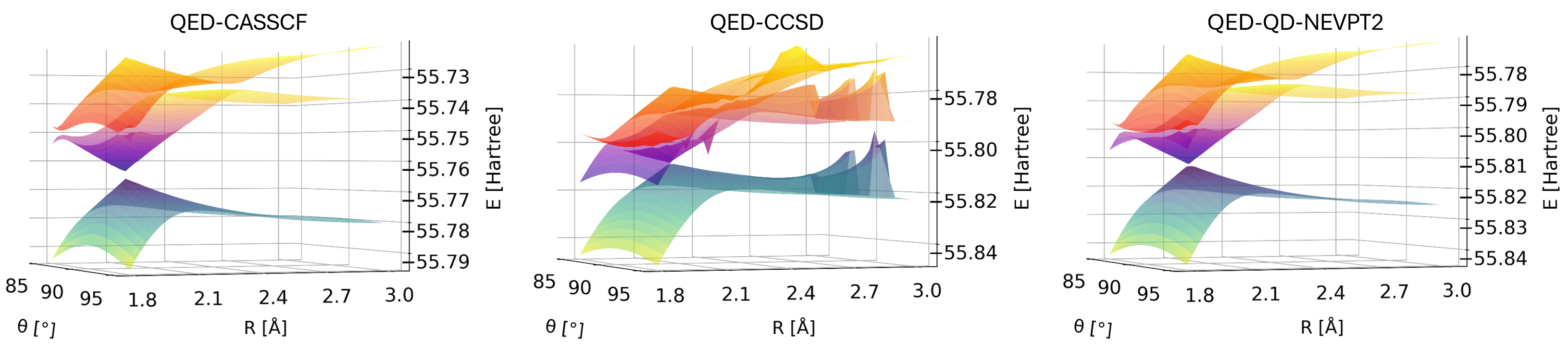}
    \caption{Polaritonic potential energy curves for the three lowest singlet states of ammonia along the $\theta$ and $R$ variables computed at the QED-CASSCF(8,7) (left), QED-CCSD (center), and QED-QD-NEVPT2 (right) levels of theory.}
    \label{fig:PES_QED_3D}
\end{figure*}
QED-CASSCF reproduces the conical intersection and the Rabi splitting qualitatively, but not quantitatively. 
Using this approach, the field seems to lift, differently from the other methods, the degeneracy between $S_0$ and $S_1$ generating an avoided crossing. This behavior is similar to the one already observed in Ref.~\citenum{haugland2020coupled} for pyrrole.

Neglecting dynamic correlation causes a shift in all QED-CASSCF curves of about 50-60 mHa compared to QED-FCI. 
The conical intersection is found at 1.975 \AA{} slightly before the QED-FCI value. 
QED-CCSD also describes the intersection correctly but, on the other side, describes the conical intersection at longer distances compared to the FCI reference and the quality of the curves deteriorates at longer bond lengths. 
Discontinuities appear at about 2.5~\AA{}, and beyond 2.75~\AA{} the curves deviate significantly from QED-FCI.
This unphysical behavior is due to the intrinsically single-reference nature of the QED-CC approach.

At the QED-QD-NEVPT2 level, the three curves run parallel to QED-FCI over the whole geometry range, with a deviation of about 8~mHa that remains essentially constant across all states (see Fig.~S2). 
Both the conical intersection and the Rabi splitting are recovered quantitatively, at the same bond lengths as in QED-FCI. 
The only visible irregularity is a local increase in the error between 1.97 and 2.0~\AA{}, caused by a near degeneracy with a fourth state that is not part of the three-state QD treatment.
This irregularity is removed when the fourth state is included (see Fig.~S3).

We also extended the scan to out-of-plane angles between 85$^{\circ}$ and 95$^{\circ}$ (Fig.~\ref{fig:PES_QED_3D}). 
QED-QD-NEVPT2 and QED-CASSCF give smooth surfaces for all three states, whereas QED-CCSD deteriorates: discontinuities appear in the excited states as well, and in some regions the surfaces cross each other. 
The same analysis in the purely electronic case (Fig.~S4) shows that CASSCF and QD-NEVPT2 give smooth surfaces with the correct shape, while the CCSD surfaces are distorted, showing discontinuities in several regions of the potential energy surfaces.
QED-CCSD therefore performs even worse than its electronic counterpart.

Overall, QED-QD-NEVPT2 provides an accurate and reliable description of the polaritonic surfaces across all the explored regions.
However, it is important to highlight that for this system and using this basis set the NEVPT2 perturbative treatment need only to fill a small gap compared to FCI. 
A more reliable analysis of the accuracy of the approach can be done on bigger systems with larger basis sets. 
This aspect is investigated in the following sections.

\subsection{\label{sec:results_etilene} Ethylene}
\begin{figure}[ht!]
    \centering
    \includegraphics[width=0.8\linewidth]{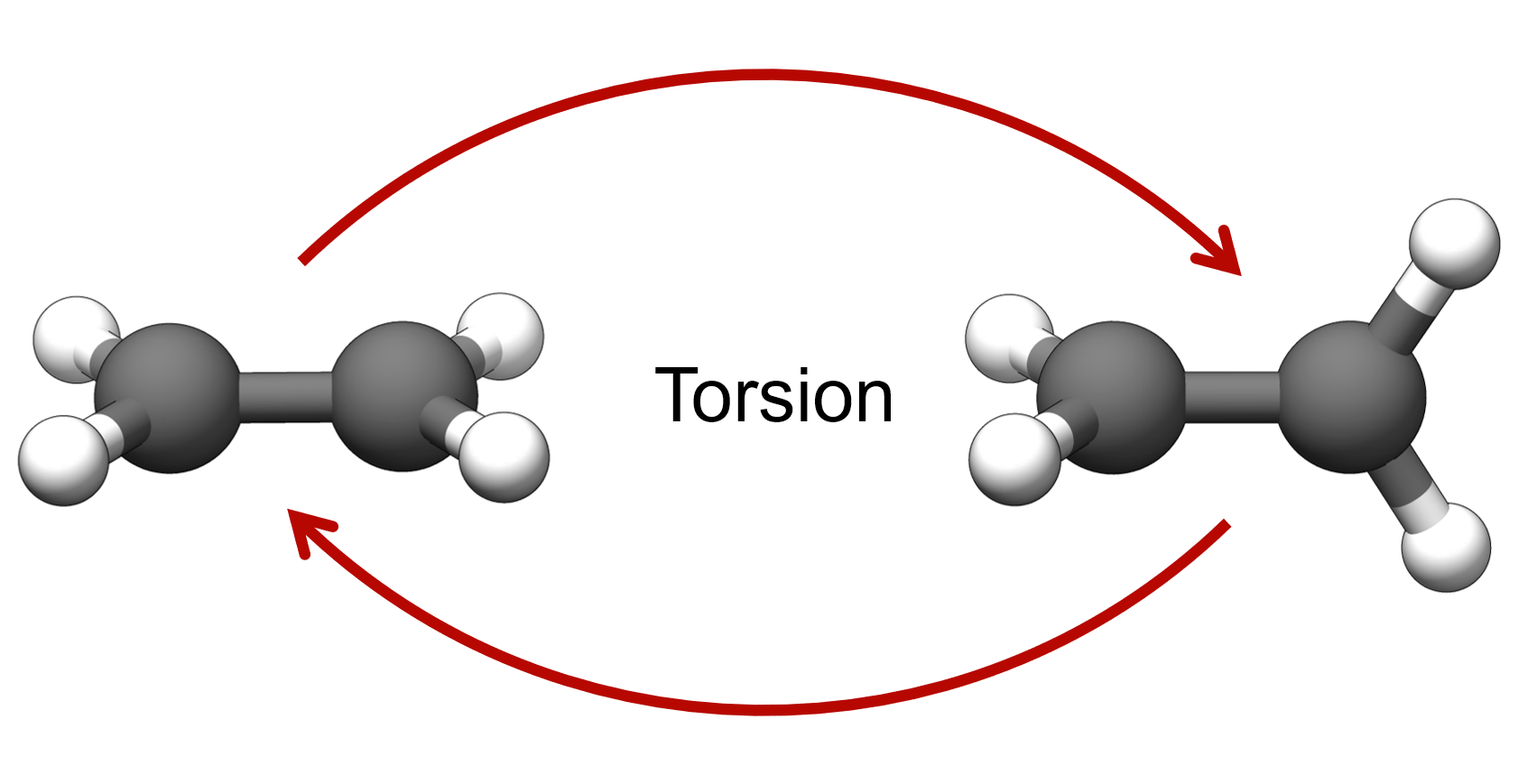}
    \caption{Representation of the ethylene torsion}
    \label{fig:Ethylene_Fig1_Torsion}
\end{figure}

In this section, the field-induced effects on the double-bond torsion of the ethylene molecule (Fig.~\ref{fig:Ethylene_Fig1_Torsion}) have been analyzed. This is a notorious multireference system commonly applied in quantum chemistry.
Starting from the planar equilibrium geometry, we scan the torsional angle $\phi$ from $0^\circ$ to $180^\circ$, keeping the rest of the molecule rigid. At $\phi = 90^\circ$ the $\pi$ bond is fully broken and the molecule acquires diradical character, bringing $S_0$ and $S_1$ close in energy~\cite{salem1972electronic, slipchenko2002singlet, barbatti2004photochemistry, zen2014static}. After this point the $\pi$ bond is restored with an inverted phase, so the two configurations at $\phi=0^\circ$ and $\phi=180^\circ$ are planar. 
We study this system using a CAS(4,4) active space in combination with the cc-pVDZ basis set. 
We first consider the three lowest singlet states at the NEVPT2 level and the four lowest singlet states at the QED-QD-NEVPT2 level. The cavity frequency is set in resonance with the $S_0 \rightarrow S_1$ electronic transition at $\phi = 40^\circ$ (or, equivalently, $\phi = 140^\circ$).

\begin{figure}[ht!]
    \centering
    \includegraphics[width=0.8\linewidth]{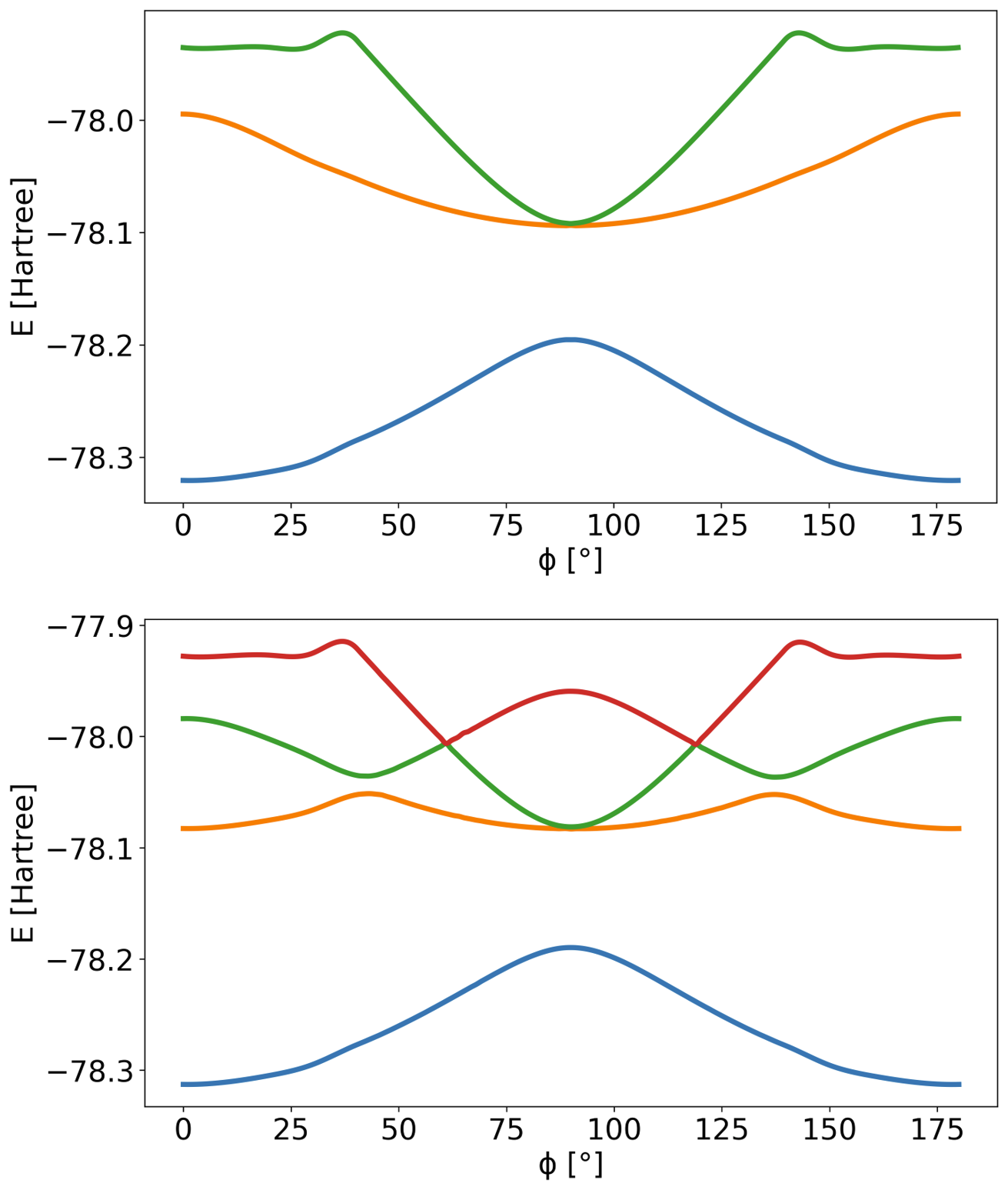}
    \caption{Potential energy curves for the three lowest singlet states of ethylene along the $\phi$ computed at NEVPT2 level (top). Polaritonic energy curves for the four lowest singlet states of ethylene along the $\phi$ computed at QED-QD-NEVPT2 level (bottom).}
    \label{fig:Ethylene_Fig2_QD_QED_NEVPT2_PES}
\end{figure}

As we can see from Fig.~\ref{fig:Ethylene_Fig2_QD_QED_NEVPT2_PES} (upper panel), electronic NEVPT2 is able to describe the curves smoothly across the entire scan, showing any near-degeneracy instabilities at $\phi = 90^\circ$.
QED-QD-NEVPT2 (Fig.~\ref{fig:Ethylene_Fig2_QD_QED_NEVPT2_PES}, lower panel), instead, correctly reproduces the Rabi splitting at $\phi = 40^\circ$ and $140^\circ$, induced by the cavity.
It is important to highlight that state-specific QED-NEVPT2 (see Fig.~S5) well captures the Rabi splitting and the crossing at $90^\circ$, but at $\phi \approx 60^\circ$ and $120^\circ$ it returns artifacts introduced by the degeneracy between the second electronic excited state and the ground state photonic replica. These observations clearly show that the coupling to the cavity field often introduces additional degeneracy point requiring a QD treatment also in cases when the standard state-specific formulation well describe the bare electronic system.
\subsection{\label{sec:results_formaldimine} Formaldimine}
As a last system, we investigate the isomerization of formaldimine (CH$_2$NH), a minimal model system widely used to study conical intersections between the ground and first excited state~\cite{yalouz2021state}. 

\begin{figure}[ht!]
    \centering
    \includegraphics[width=0.75\linewidth]{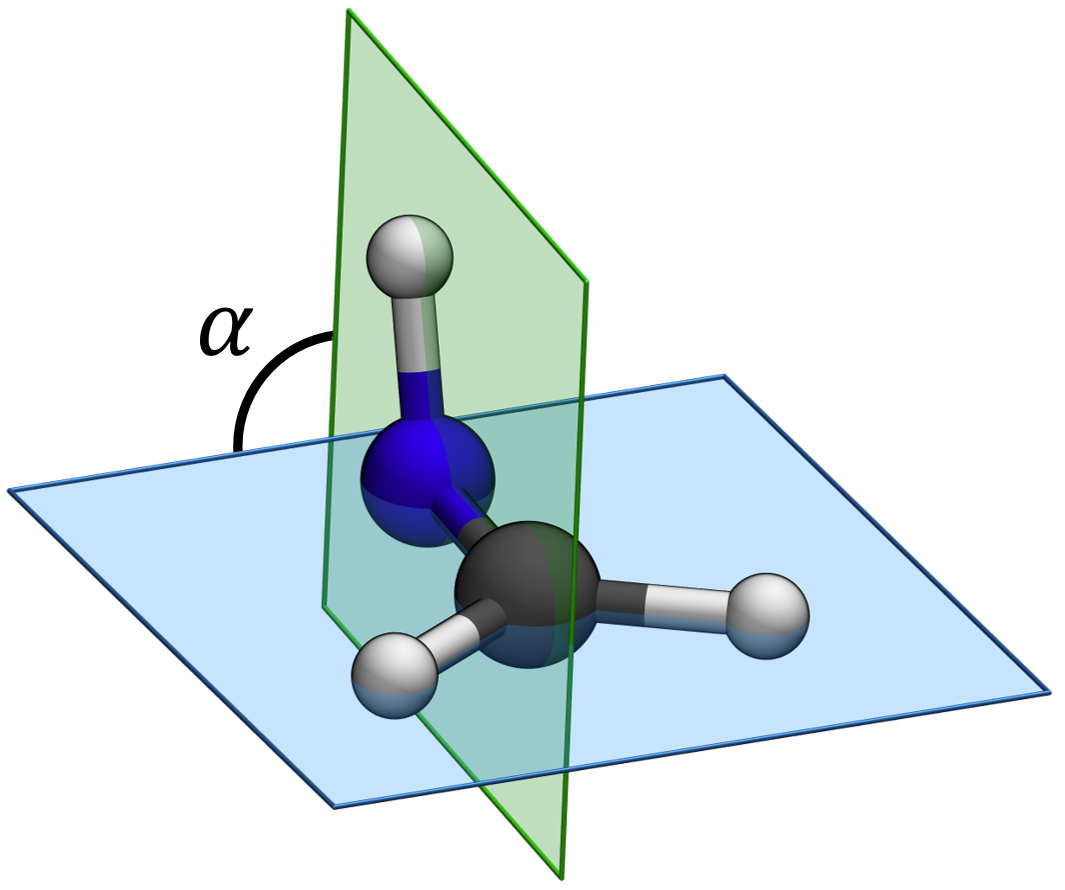}
    \caption{Representation of the dihedral angle in formaldimine}
    \label{fig:Formaldimine_Fig1_Molecule}
\end{figure}

In this work we focus, instead, on the effects generated by the field on a scan of the dihedral angle $\alpha = \text{H--C--N--H}$ (Fig.~\ref{fig:Formaldimine_Fig1_Molecule}) while keeping the remaining coordinates frozen.
We employ a (4,4) active space, corresponding to the $\sigma$ and $\pi$ orbitals in combination with the cc-pVDZ basis set. 
The cavity mode is tuned in resonance with the $S_0 \rightarrow S_1$ transition at $\alpha = 50^\circ$ (and equivalently at $\alpha = 130^\circ$).
\begin{figure}[h!]
    \centering
    \includegraphics[width=0.9\linewidth]{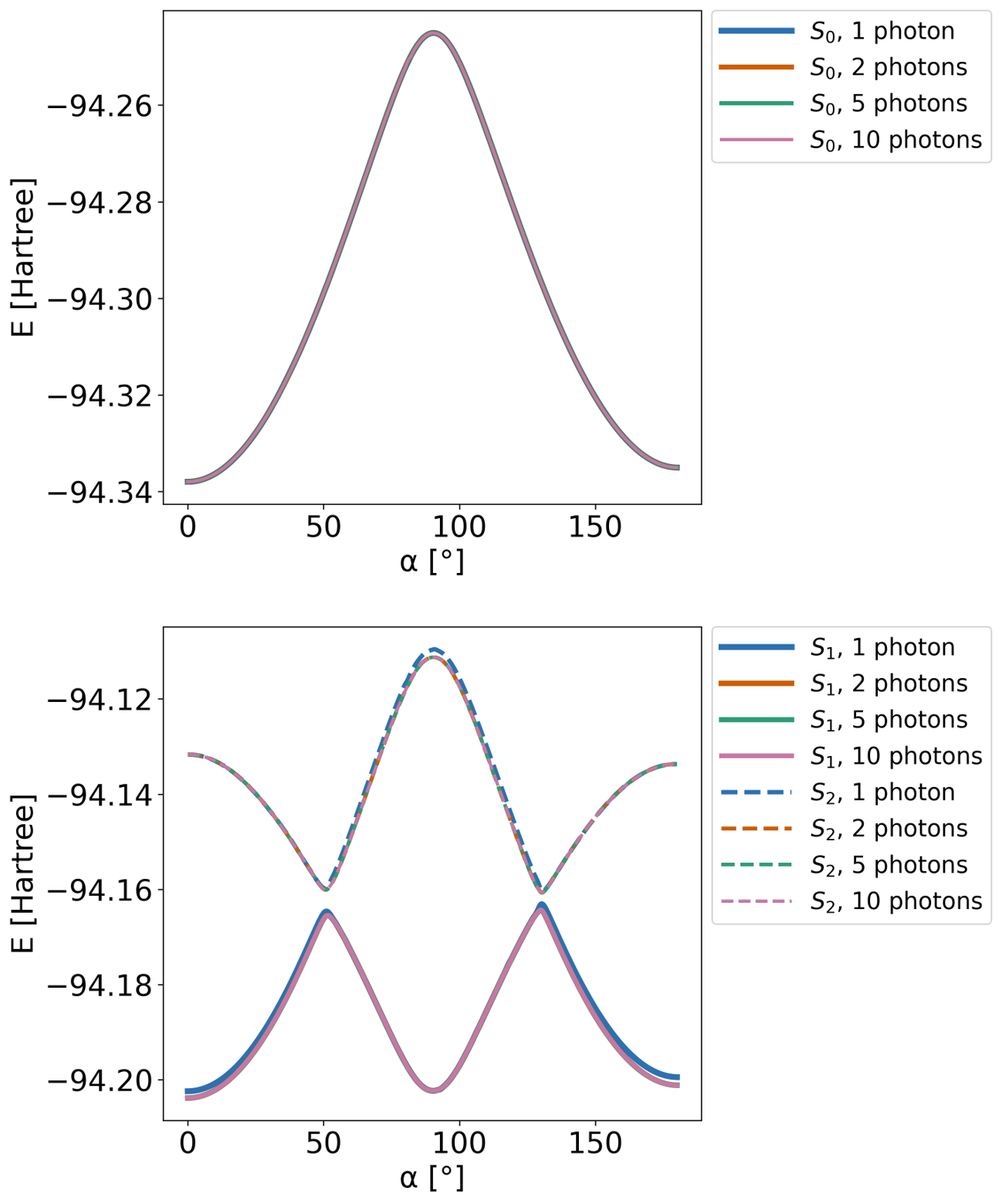}
    \caption{QED-NEVPT2 potential energy curves along $\alpha$ computed with 1, 2, 5, and 10 photons. $S_0$ curve is shown in the top panel, the $S_1$ and $S_2$ curves in the bottom panel.}
    \label{fig:Formaldimine_Fig2_QED_NEVPT2_S0S1S2}
\end{figure} 
Unlike the ammonia and ethylene cases discussed before, QED-NEVPT2 (Fig.~\ref{fig:Formaldimine_Fig2_QED_NEVPT2_S0S1S2}) correctly reproduces the Rabi splitting between $S_1$ and $S_2$ at the two resonance angles.
This behavior can be explained by the absence of conical intersections in the potential energy surfaces of formaldimine. 
In this case indeed, the non-zero transition dipole moment between the ground state photonic replica and the excited state generates the Rabi-splitting (avoided crossing). This generate a gap between $S_1$ and $S_2$ that consequently are not affected by degeneracy effects as those observed in the previous systems.
Looking at the top panel of Fig.~\ref{fig:Formaldimine_Fig2_QED_NEVPT2_S0S1S2} the $S_0$ curve is practically unaffected by the coupling and, for this coupling value, is already converged with a single photon state. On the other hand, the $S_1$ and $S_2$ curves need at least two photonic states to reach convergence. For this reason, the following results have been obtained by including two photonic states.
\begin{figure*}[ht!]
    \centering
    \includegraphics[width=\linewidth]{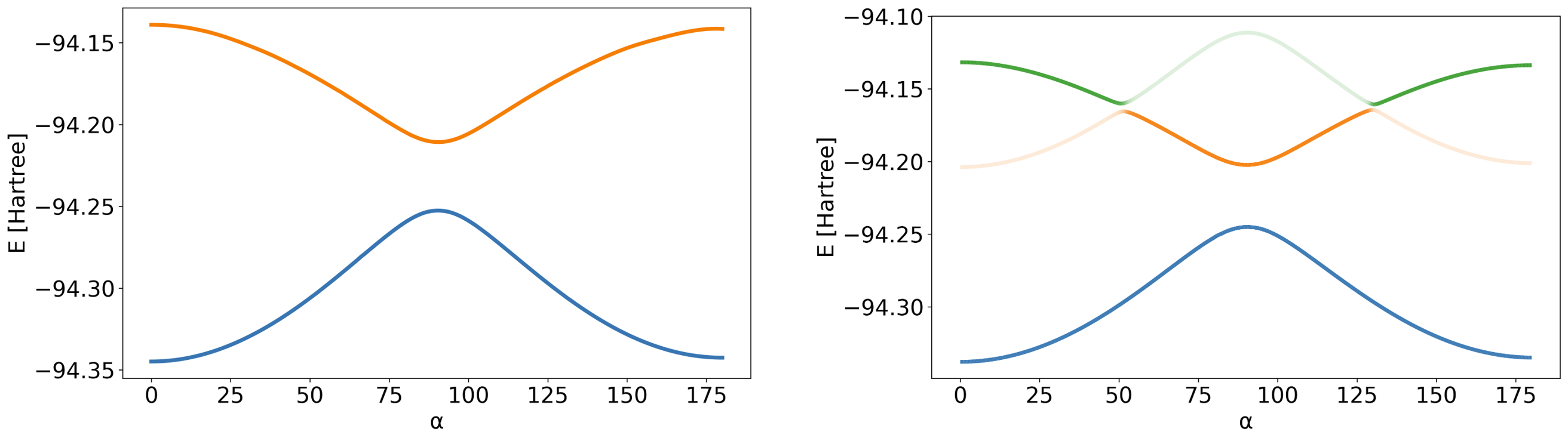}
    \caption{NEVPT2 potential energy curves of the two lowest electronic states (left). QED-NEVPT2 curves of the three lowest polaritonic states (right). The more transparent a curve, the higher its photonic character.
    }
    \label{fig:Formaldimine_Fig3_NEVPT2_vs_QED_NEVPT2_2Ph}
\end{figure*} 
Figure~\ref{fig:Formaldimine_Fig3_NEVPT2_vs_QED_NEVPT2_2Ph} compares the purely electronic NEVPT2 curves with the QED-NEVPT2 ones, and shows the photonic character of each polaritonic state. $S_0$ is essentially identical to its electronic counterpart throughout the scan.  Away from the resonance (for $\alpha \lesssim 50^\circ$ and $\alpha \gtrsim 130^\circ$), $S_1$ is dominated by the photonic replica of $S_0$ while $S_2$ is mainly described by the first electronic excited state. The two curves cross at $\alpha \approx 50^\circ$ and $130^\circ$, exchange their character in correspondence of the Rabi splitting, and between the two angles the order is reversed: $S_1$ gain electronic character, and is mainly characterized by the electronic excited state, and $S_2$ gain photonic character, and is dominated by the electronic $S_0$ with an additional photon.

\begin{figure*}[t]
    \centering
    \includegraphics[width=\linewidth]{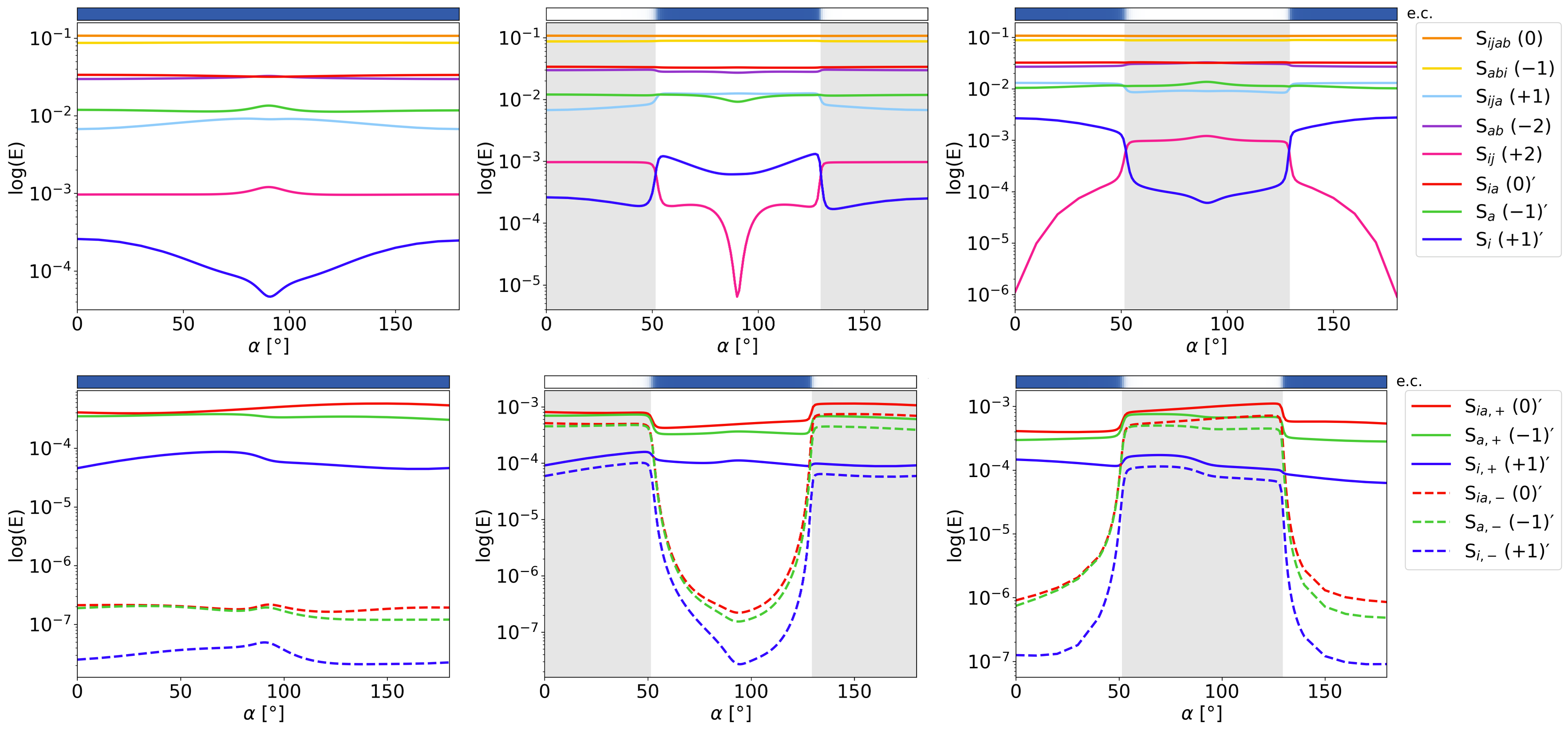}
    \caption{Contributions of the individual perturber classes to the QED-NEVPT2 second-order energy, in logarithmic scale, as a function of the angle $\alpha$. The columns refer to $S_0$, $S_1$ and $S_2$ (left to right). The top row shows the purely electronic perturbers $\ket{K}$ and the bottom row the mixed perturbers ($\ket{K_+}$ solid, $\ket{K_-}$ dashed). The bar above each plot gives the electronic character of the state: the bluer, the more electronic.}
    \label{fig:Formaldimine_Fig5_Perturbers_SOS1S2}
\end{figure*}

In Fig.~\ref{fig:Formaldimine_Fig5_Perturbers_SOS1S2} we analyze the perturbers contributions to the total second-order energy.

Starting from the electronic perturbers, the subspaces $S_{ijab}^{(0)}$ and $S_{abi}^{(-1)}$ dominate by about two orders of magnitude and the subspaces $S_{i}^{(+1)'}$ and $S_{ij}^{(+2)}$ are those that contribute less to the total energy correction. In the $S_1$ and $S_2$ curves, sharp changes are displayed at the two resonance angles where the states exchange character, although the perturber classes $\ket{K}$ exhibit a smooth and uniform behavior. These abrupt changes are not divergences. A vanishing denominator would produce a spike in the contribution of the corresponding class, whereas all the contributions in Fig.~\ref{fig:Formaldimine_Fig5_Perturbers_SOS1S2} remain bounded and smooth along the whole scan, and the steps at the two resonance angles follow the exchange of electronic and photonic character of the states. 

The mixed perturbers are significantly smaller and are much more sensitive to the photonic content of the state.
In $S_0$, which has a very small photonic character, $\ket{K_+}$ is on the order of $10^{-4}$ while $\ket{K_-}$ is three to four orders of magnitude smaller. This is reasonable, since acting with a photonic annihilation operator on an almost pure electronic state produces negligible corrections to the energy.
In $S_1$ and $S_2$ curves, in the regions where the electronic contribution dominate, the same behavior is followed.

For these latter states, in regions where the photonic character dominates, the $\ket{K_-}$ contributions rise by several orders of magnitude and become comparable to the $\ket{K_+}$ ones. For $S_1$ this happens for $\alpha \lesssim 50^\circ$ and $\alpha \gtrsim 130^\circ$, for $S_2$ instead it happens in between, as expected from the character exchange at the Rabi splitting. The $\ket{K_-}$ perturbers can therefore be interpreted as a ``marker'' of the photonic character of a polaritonic state even if they are small in absolute terms.

\begin{figure}[ht!]
    \centering
    \includegraphics[width=\linewidth]{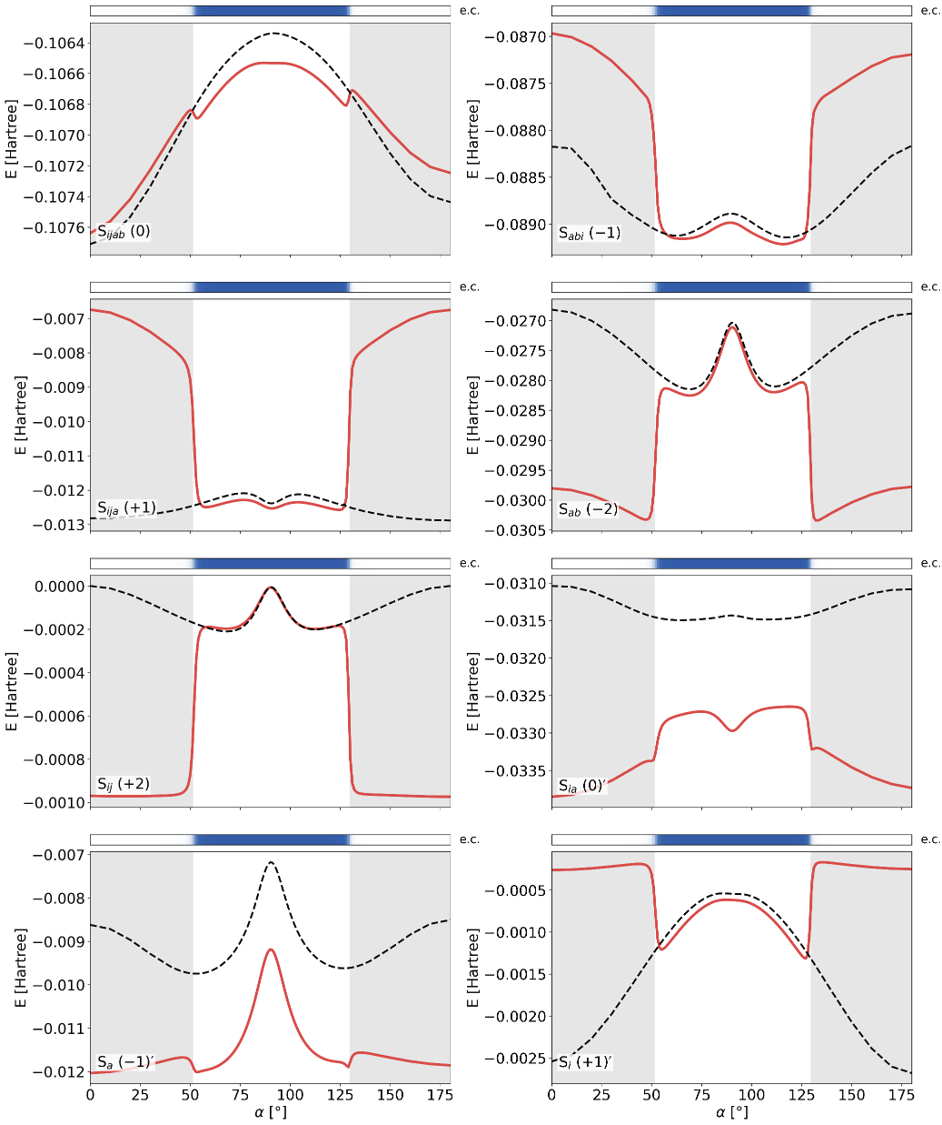}
    \caption{Contribution of each electronic perturber class to the second-order energy of $S_1$ as a function of $\alpha$. QED-NEVPT2 is represented with red solid lines and NEVPT2 with black dashed lines. The bar above each panel gives the electronic character of the QED state and the gray regions mark PES sector dominated by photonic contributions.}
    \label{fig:Formaldimine_Fig5_Perturbers_S1_NEVPT2_vs_QED_NEVPT2}
\end{figure}

Finally, we compare each electronic perturber class with its QED counterpart for $S_1$ (Fig.~\ref{fig:Formaldimine_Fig5_Perturbers_S1_NEVPT2_vs_QED_NEVPT2}). The same comparison for $S_0$ is reported in the Supporting Information (Fig.~S6).
For $S_0$, which is purely electronic along the whole scan, the QED and electronic contributions have the same shape and differ only by a small shift. For $S_1$ the two sets of curves are close where the state is mostly electronic for almost all perturbers. On the external gray regions, where the photon character increases, the QED contributions significantly diverge from the electronic ones, in size and shape. 
\section{\label{sec:conclusions} Concluding Remarks}
We have extended second-order N-electron valence state perturbation theory to polaritonic systems, in its state-specific (QED-NEVPT2) and quasidegenerate (QED-QD-NEVPT2) formulations. 
Starting from a QED-CASSCF reference, which incorporates the correlation associated with the presence of more degenerate configurations, the methods add the remaining correlation coming from instantaneous (dynamical) interactions between the involved quantum particles. 
Two new families of perturbers, $\ket{K_+}$ and $\ket{K_-}$, which are composed by a single electronic excitation and a creation or annihilation of a photon have been introduced. Following the spin-free commutator formalism of Angeli and co-workers~\cite{Angeli2002}, the energy corrections have been derived. 
The applications of these methodologies to the ammonia, ethylene and formaldimine molecules show that the use of the quasidegerate treatment of the approach becomes compulsory when degeneracies between different states come into play.
In polaritonic systems, the appearance of degeneracy situations is more likely; this is due to the crossing between electronic states and photonic replicas induced by the field. 
It is important to highlight that, in these conditions, the QED Coupled Cluster approach -- the reference \textit{ab-initio} method for polaritonic systems -- is dramatically failing (as demonstrated on ammonia in this work) in the description of polaritonic potential energy surfaces.
These limitations are very severe if dissociated molecular systems need to be investigated. This clearly explains why QED-CC has not been actively used so far to investigate cavity induced effects on the ground state reactivity and photochemistry of molecular systems; the most debated aspect of polaritonic chemistry since a decade.
For all these reasons, QED-NEVPT2 has all the right characteristics to become the next reference methodology to investigate polaritonic molecular systems in particular when strong multi-configurational effects become crucial.

The method is currently limited to a single cavity mode in the dipole approximation.
However, it can be extended to the multi-mode case and to account for the explicit spatial dependence of the electromagnetic field~\cite{riso2023strong} in a relatively easy manner. 
In the near future, we also plan to couple it with approximate treatments to include collective effects~\cite{castagnola2025realistic} due to the presence of the other molecules or the solvent. Approximate strategies to treat cavity dissipation effects can also be introduced~\cite{bruno2026beyond}.

Concluding, we proposed the first \textit{ab-initio} approach for treating in a complete manner the electron-electron and electron-photons correlation in polaritonic multi-configurational molecular systems. We believe that this approach will represent a major step forward toward a realistic simulation of reactive events in optical cavities. 

\begin{acknowledgments}
E.A., R.A., and E.R. acknowledge funding from the European Research Council (ERC) under the European Union’s Horizon Europe Research and Innovation Programme (Grant n. ERC-StG-2021-101040197 - QED-SPIN).
The authors are thankful to Federico Rossi for providing the geometries for the ammonia scan and to Saad Yalouz and Henrik Koch for insightful discussions.
\end{acknowledgments}

\section*{Data availability}
The QED-CCSD and QED-FCI calculations can be performed with the development version of the $e^{\textit{T}}$ program available at the following link:~\href{https://gitlab.com/eT-program/eT}{gitlab.com/eT-program/eT}.
The $e^{\textit{T}}$ program used to perform the QED-(QD-)NEVPT2 calculations shown in this work is available from the corresponding authors upon
reasonable request. 
Scripts to generate and run inputs as well as the \texttt{.blend} file to generate the 3D image in Fig.~\ref{fig:Cartoon} are available at the following Zenodo repository:~\href{https://zenodo.org/records/23503732}{zenodo.org/records/23503732}.

\newpage
\appendix
\begin{widetext}
\section{Electronic perturbers}\label{AppendixA}
This appendix reports the contributions to the second-order energy of the $S_l^{(k)}$ subspaces. In order to provide some details about the notation, we follow the formalism of Ref.~\cite{Angeli2002}, where $R^{(n)}$ and $\Tilde{R}^{(n)}$ denote the $n$-body reduced density matrix and the $n$-body hole reduced density matrix  with all indices restricted to the active orbitals. The bilinear coupling introduces terms in which the electronic excitation operators are multiplied by $(b^{\dagger}+b)$. For readability we define the one-body bilinear density matrix
\begin{equation}
    \begin{split}    
    \gamma_{pq} &= \sum_{\Phi,\Theta} \sum_m \Bigg[ C_{\Phi,m+1} \matrixel{\Phi}{E_{pq}}{\Theta}
    C_{\Theta,m} \sqrt{m+1} \\
    &+ C_{\Phi,m-1} \matrixel{\Phi}{E_{pq}}{\Theta}
    C_{\Theta,m} \sqrt{m} \Bigg]
    \end{split}
\end{equation}
and its hole counterpart $\Tilde{\gamma}_{pq}$,
\begin{equation}
    \begin{split}    
    \Tilde{\gamma}_{pq} &= \sum_{\Phi,\Theta} \sum_m \Bigg[ C_{\Phi,m+1} \matrixel{\Phi}{\Tilde{E}_{pq}}{\Theta}
    C_{\Theta,m} \sqrt{m+1} \\
    &+ C_{\Phi,m-1} \matrixel{\Phi}{\Tilde{E}_{pq}}{\Theta}
    C_{\Theta,m} \sqrt{m} \Bigg]
    \end{split}
\end{equation}
Similarly, the two-body bilinear density matrix and its hole counterpart are
\begin{equation}
    \begin{split}
    \Gamma_{pqrs} &= \sum_{\Phi,\Theta} \sum_m \Bigg[C_{\Phi,m+1} \matrixel{\Phi}{a^{\dag}_p a^{\dag}_q a_r a_s }{\Theta}C_{\Theta,m} \sqrt{m+1}\\
    &+ C_{\Phi,m-1} \matrixel{\Phi}{a^{\dag}_p a^{\dag}_q a_r a_s }{\Theta}C_{\Theta,m} \sqrt{m} \Bigg]
    \end{split}
\end{equation}
\begin{equation}
    \begin{split}
    \Tilde{\Gamma}_{pqrs} &= \sum_{\Phi,\Theta} \sum_m \Bigg[ C_{\Phi,m+1} \matrixel{\Phi}{ a_p a_q a^{\dag}_r a^{\dag}_s }{\Theta}C_{\Theta,m} \sqrt{m+1} \\
    &+ C_{\Phi,m-1} \matrixel{\Phi}{a_p a_q a^{\dag}_r a^{\dag}_s}{\Theta}C_{\Theta,m} \sqrt{m} \Bigg]
    \end{split}
\end{equation}
\subsection{Subspaces $\mathbf{S_{ij,ab}^{(0)}}$}
For this subspace the norm is defined as 
\begin{equation}
    N^{(0)}_{ij,ab} = \eta_{ab} \eta_{ij} (aj||bi)^{2}
\end{equation}
The contribution to the second-order energy is given  by the following M\o{}ller--Plesset (MP)-like term:
\begin{equation}
    E^{(2)}_{I} (S_{ij,ab}^{(0)})= - \frac{N^{(0)}_{ij,ab}}{\Delta\epsilon_{ij,ab}^{(0)}}
\end{equation}
where
\begin{equation}
    \begin{split}
    \Delta\epsilon_{ij,ab}^{(0)} &= \epsilon_a + \epsilon_b - \epsilon_i - \epsilon_j \\
    &+ \sqrt{\frac{\omega}{2}} \Bigg[\sum_{\Phi} \sum_{m} \left(d_{aa} + d_{bb} - d_{jj} - d_{ii}\right)  \\
    & \times \bigg(C_{\Phi,m} C_{\Phi,m+1} \sqrt{m+1}
    +  C_{\Phi,m} C_{\Phi,m-1} \sqrt{m}\bigg)  \Bigg]
    \end{split}
\end{equation}
\subsection{Subspaces $\mathbf{S_{ab,i}^{(-1)}}$}
For this subspace the norm is defined as 
\begin{equation}
    N^{(-1)}_{ab,i}
    =  \eta_{ab}\sum_{u'u} (au'||bi) (au||bi) R^{(1)}_{u'u}
\end{equation}
The contribution arising from  $H_i$ is 
\begin{equation}
    \begin{split}    
    \Delta\epsilon_{ab,i}^{(-1)} &= \epsilon_a + \epsilon_b - \epsilon_i + \eta_{ab} \sqrt{\frac{\omega}{2}} \sum_{u'u}\Bigg[  (au'||bi) (au||bi) \\
    &\times \big( d_{bb} + d_{aa} - d_{ii} \big) \gamma_{u'u} \Bigg] / N^{(-1)}_{ab,i}
    \end{split}
\end{equation}
The $H_v$ contribution is 
\begin{equation}
         \xi_{ab,i}^{(-1)}= \frac{1}{N^{(-1)}_{ab,i}} \eta_{ab} \sum_{u,u'} 
        (au'||bi) (au||bi) K_{u'u}
\end{equation}
with 
\begin{equation}\label{eq:koopmans-1}
    \begin{split}
        K_{u'u} &= \frac{1}{2}\matrixel{\Psi^{(0)}_{I}}{E_{u'b}E_{ia} [H_{v}, E_{ai}E_{bu}]}{\Psi^{(0)}_{I}} \\
        &= - \sum_{w} h_{uw} ^{eff} R_{u'w} ^{(1)}
        - \sum_{wxy} g_{uwxy} R^{(2)}_{u'wxy} \\
        &- \sqrt{\frac{\omega}{2}} \sum_{v} d_{uv} \gamma_{u'v}
    \end{split}
\end{equation}
$K_{u'u}$ is the QED equivalent of Koopmans' matrix but it is important to note that due to the different nature of $ H_v$, $K_{u'u}$ cannot be associated to any physical phenomena within the active space. 
Therefore, the contribution to the second-order energy is:
\begin{equation}
    E^{(2)}_{I}(S_{ab,i}^{(-1)})=- \frac{N^{(-1)}_{ab,i}}{\Delta\epsilon^{(-1)}_{ab,i}+ \xi_{ab,i}^{(-1)}} 
\end{equation}
\subsection{Subspaces $\mathbf{S^{(+1)}_{ij,a}}$}
The evaluation of energy contribution of for $S_{ij,a}^{(+1)}$ is very similar to the previous one. In this case, the norm is defined as
\begin{equation}
    N^{(+1)}_{ij,a} =  \eta_{ij} \sum_{u'u}(aj||u'i) (aj||ui) \Tilde{R}^{(1)}_{u'u}
\end{equation}
The contribution arising from  $H_i$ is 
\begin{equation}
    \begin{split}
    \Delta\epsilon^{(+1)}_{ij,a}&= \epsilon_a - \epsilon_i - \epsilon_j + \eta_{ij} \sqrt{\frac{\omega}{2}}  \sum_{u'u}\Bigg[   (aj||u'i) (aj||ui)\\
    & \times \left( d_{aa} - d_{ii} - d_{jj}\right) \Tilde{\gamma}_{u'u} \Bigg]/N^{(+1)}_{ij,a}
    \end{split}
\end{equation}
The active contribution is 
\begin{equation}
    \xi^{(+1)}_{ij,a}= \frac{1}{N^{(+1)}_{ij,a}} \eta_{ij}\sum_{u,u'} (aj||u'i) (aj||ui) K' _{u'u}
\end{equation}
with
\begin{equation}\label{eq:koopmans+1}
    \begin{split}
    K' _{u'u} &= \frac{1}{2}\matrixel{\Psi^{(0)}_{I}}{E_{ju'}E_{ia} [H_{v}, E_{ai}E_{uj}]}{\Psi^{(0)}_{I}} \\
    &= \sum_{v} h_{vu}^{eff} \Tilde{R}^{(1)}_{u'v} + \sum_{vxy} g_{uvxy} 
                \matrixel{\Psi^{(0)}_{I}}{\Tilde{E}_{u'v}E_{xy}}{\Psi^{(0)}_{I}} \\
    &  + \sqrt{\frac{\omega}{2}} \sum_{v} d_{vu} \Tilde{\gamma}_{u'v}
    \end{split}
\end{equation}
The $K'_{u'u}$ matrix exhibits a structure similar to a Koopmans' electron affinity matrix, augmented by an additional active bilinear term. The energy contribution to the second-order energy is given by:
\begin{equation}
    E^{(2)}_{I}(S^{(+1)}_{ij,a})= - \frac{N^{(+1)}_{ij,a}}{\Delta\epsilon^{(+1)}_{ij,a} + \xi^{(+1)}_{ij,a}}
\end{equation}
\subsection{Subspace $\mathbf{S^{(-2)}_{ab}}$}
For this subspace the norm is defined as 
\begin{equation}
    N^{(-2)}_{ab}= \eta_{ab}
    \sum_{u'v'uv} (au'||bv') (au||bv) R_{v'u'uv}^{(2)}
\end{equation}
The contribution arising from  $H_i$ is
\begin{equation}
    \begin{split}
    \Delta\epsilon^{(-2)}_{ab} &= \epsilon_b + \epsilon_a + \eta_{ab}
     \sqrt{\frac{\omega}{2}} \sum_{u'v'uv} \Bigg[  (au'||bv') (au||bv) \\
    &\times  \left( d_{aa} + d_{bb} \right) \Gamma_{v'u'uv} \Bigg]/N^{(-2)}_{ab}
    \end{split}
\end{equation}
The contribution arising from  $H_v$ is 
\begin{equation}
    \xi^{(-2)}_{ab} = \frac{1}{N^{(-2)}_{ab}} \eta_{ab}\sum_{u'v'u v} 
    ( au | bv)( au' | bv')K_{v'u'uv}
\end{equation}
where
\begin{equation}\label{eq:koopmans-2}
    \begin{split}
    K_{v'u'vu} &= \matrixel{\Psi^{(0)}_{I}}{E_{v'b}E_{u'a} [H_{v}, E_{au}E_{bv}]}{\Psi^{(0)}_{I}} \\   
    &= - \sum_{x} \left( h_{vx} ^{eff'} R^{(2)}_{v'u'ux} + h_{ux} ^{eff'} R^{(2)}_{v'u'xv}
    \right) \\
    & - \frac{1}{2} \sum_{xyz} \Big[
    g_{vxyz} \bigg(2R^{(3)}_{v'u'y,zux} \\
    &+ R^{(2)}_{v'u',yx}\delta_{uz}                         
    + R^{(2)}_{v'u',zu}\delta_{xy} \bigg)\\
    & + g_{uxyz} \left( 2R^{(3)}_{v'u'y,vzx} + R^{(2)}_{v'u',xz}\delta_{yv}                            + R^{(2)}_{v'u',vz}\delta_{xy}\right)\Big]\\
    & -\sqrt{\frac{\omega}{2}} \Big[
    \sum_{w} \left(d_{vw} \Gamma _{v'u'uw} + d_{uw} \Gamma _{v'u'wv} \right)\Big] 
    \end{split}
\end{equation}
The $K_{v'u'uv}$ matrix exhibits a structure similar to a Koopmans' double ionization matrix, augmented by an additional active bilinear term. The contribution to the second-order energy is given by:
\begin{equation}
    E^{(2)}_{I}(S^{(-2)}_{ab})= - \frac{N^{(-2)}_{ab}}{\Delta\epsilon^{(-2)}_{ab} + \xi^{(-2)}_{ab}}
\end{equation}
\subsection{Subspace $\mathbf{S_{ij}^{(+2)}}$}
This case closely follows the previous one and the norm is defined as 
\begin{equation}
    N^{(+2)}_{ij} = \eta_{ij} \sum_{u'v'uv} (u'i||v'j) ( ui||vj) \Tilde{R}_{v' u'uv}^{(2)}
\end{equation}
The contribution arising from  $H_i$ is 
\begin{equation}
    \begin{split}
    \Delta\epsilon_{ij}^{(+2)} &= - \epsilon_{j} - \epsilon_{i} - \eta_{ij}  \sqrt{\frac{\omega}{2}}  \sum_{u'v'uv} \Bigg[  (u'i||v'j) ( ui||vj)  \\
    &\times (d_{jj} + d_{ii}) \Tilde{\Gamma}_{v'u',uv} \Bigg] /N^{(+2)}_{ij}
    \end{split}
\end{equation}
The contribution arising from  $H_v$ is 

\begin{equation}
    \xi_{ij}^{(+2)} =  \frac{1}{N^{(+2)}_{ij}} \eta_{ij}\sum_{u'v'u v} 
    ( ui | vj)( u'i | v'j) K'_{v'u'uv}
\end{equation}
where
\begin{equation}\label{eq:koopmans+2}
    \begin{split}
        K'_{v'u'uv}&=\matrixel{\Psi^{(0)}_{I}}{E_{jv'}E_{iu'} [H_{v}, E_{ui}E_{vj}]}{\Psi^{(0)}_{I}}\\
        &= \sum_{w} \left( h^{eff'}_{wv} \Tilde{R}^{(2)}_{v'u',uw} + h^{eff'}_{wu}
        \Tilde{R}^{(2)}_{v'u',wv} \right)\\
        & + \frac{1}{2}\sum_{wxy} \Big[ g_{wxyu}\bigg( - 2 \Tilde{R}^{(3)}_{v'u'y,wxv} +4\Tilde{R}^{(2)}_{v'u',xv}\delta_{wy} \\
        &-\Tilde{R}^{(2)}_{v'u',wv}\delta_{xy} - \Tilde{R}^{(2)}_{v'u',xw}\delta_{vy}\bigg)\\
        &+g_{wxyv}\bigg( - 2 \Tilde{R}^{(3)}_{v'u'y,uwx} + 4 \Tilde{R}^{(2)}_{v'u',ux}\delta_{wy} \\
        &-\Tilde{R}^{(2)}_{v'u',wx}\delta_{uy} - \Tilde{R}^{(2)}_{v'u',uw}\delta_{xy} \bigg) \Big] \\
        &+  \sqrt{\frac{\omega}{2}}  \Bigg[\sum_{w} \left(d_{wv} \Tilde{\Gamma}_{v'u',uw} + d_{wu} \Tilde{\Gamma}_{v'u',wv} \right) \Bigg]    
    \end{split}
\end{equation}
The $K'_{v'u'uv}$ matrix exhibits a structure similar to a Koopmans' double electron affinity matrix, modified with an additional active bilinear term. The energy contribution to the second-order energy is given by:
\begin{equation}
    E^{(2)}_{I}(S_{ij}^{(+2)}) = - \frac{N^{(+2)}_{ij}}{\Delta\epsilon_{ij}^{(+2)} + \xi_{ij}^{(+2)}}
\end{equation}
\subsection{Subspaces  $\mathbf{S^{(0)'}_{i,a}}$}
For this subspace, the norm is defined as 
\begin{equation}
    \begin{split}
        N^{(0)'}_{i,a} &=\sum_{u'v'uv} [ \Big(2(ai | u'v')(ai| uv) - (u'i | av')(ai | uv) - (ai | u'v')(ui | av)\Big)\matrixel{\Psi^{(0)}_{I}}
        {E_{v'u'}E_{uv}}{\Psi^{(0)}_{I}} \\    
        & + (u'i | av')(ui | av) (\matrixel{\Psi^{(0)}_{I}}{E_{v'v}\Tilde{E}_{u'u}}{\Psi^{(0)}_{I}} + \delta_{uv}R^{(1)}_{v'u'})] + 2(h^{eff}_{ai})^{2}\\
        & +\sum_{u'v'} [2(ai | u'v') -(u'i| av')] h^{eff}_{ai} R^{(1)}_{v'u'} + \sum_{uv} [2(ai | uv) -(ui| av)] h^{eff}_{ai} R^{(1)}_{uv}
    \end{split}
\end{equation}
We have introduced an intermediate matrix $\overline{h}^{eff}_{pq}$ that will frequently appear in the results from this point and is defined as:
\begin{equation}
    \overline{h}^{eff}_{pq} = {h}^{eff}_{pq} - \frac{1}{2} \sum_{u} {g}_{puuq}
\end{equation}
The contribution arising from  $H_v$ is 
\begin{equation}
    \begin{split}
    \xi_{i,a}^{(0)'}&= \frac{1}{N^{(0)'}_{i,a}} \matrixel{\Psi^{(0)}_{I}}{\left( V^{(0)'}_{i,a} \right) ^\dag \big[H_{v}, V^{(0)'}_{i,a}\big]}{\Psi^{(0)}_{I}}\\
    &=  \frac{1}{N^{(0)'}_{i,a}} \sum_{u'v'uv}
    \Bigg\{ \Big[ 2 (ai|u'v')(ai|uv) - (ai|u'v')(ui|av) - (u'i|av')(ai|uv) \Big] A_{v'u',uv} + (u'i|av')(ui|av) B_{v'u',uv}  \Bigg\}
    \end{split}
\end{equation}
where
\begin{align}
    A_{v'u',uv} &= \frac{1}{2} \matrixel{\Psi^{(0)}_{I}}{E_{v'u'}E_{ia}[H_v,E_{ai}E_{uv}]}{\Psi^{(0)}_{I}} \nonumber \\
    &= \sum_{w} \Bigg[ \overline{h}^{eff}_{wu} \matrixel{\Psi^{(0)}_{I}}{E_{v'u'}E_{wv}}{\Psi^{(0)}_{I}} -\overline{h}^{eff}_{vw}\matrixel{\Psi^{(0)}_{I}}{E_{v'u'}E_{uw}}{\Psi^{(0)}_{I}}
    \Bigg] \nonumber \\
    &+ \frac{1}{2}\sum_{wxy} \Bigg[ 
    {g}_{wxyu}\left(\matrixel{\Psi^{(0)}_{I}}{E_{v'u'}E_{yv}E_{wx}}{\Psi^{(0)}_{I}}
    + \matrixel{\Psi^{(0)}_{I}}{E_{v'u'}E_{wx}E_{yv}}{\Psi^{(0)}_{I}}\right)\nonumber \\
    &-{g}_{wxvy}\left(\matrixel{\Psi^{(0)}_{I}}{E_{v'u'}E_{uy}E_{wx}}{\Psi^{(0)}_{I}}
    +\matrixel{\Psi^{(0)}_{I}}{E_{v'u'}E_{wx}E_{uy}}{\Psi^{(0)}_{I}}\right) \Bigg]\\
    \nonumber \\
    B_{v'u',uv} &= \matrixel{\Psi^{(0)}_{I}}{E_{v'a}E_{iu'}[H_v,E_{ui}E_{av}]}{\Psi^{(0)}_{I}}  \nonumber \\
    &=  \sum_{w}\Bigg\{\Big[ -\overline{h}^{eff}_{vw} \left(\matrixel{\Psi^{(0)}_{I}}{\Tilde{E}_{u'u}E_{v'w}}{\Psi^{(0)}_{I}} + \delta_{v'u'} R^{(1)}_{uw} \right)\Big] +\Big[\overline{h}^{eff}_{wu} \left(\matrixel{\Psi^{(0)}_{I}}{\Tilde{E}_{u'w}E_{v'v}}{\Psi^{(0)}_{I}} + \delta_{v'u'}R^{(1)}_{wv} \right) \Big]
    \Bigg\}\nonumber \\
    &+ \frac{1}{2}\sum_{wxy} \Bigg\{ - g_{vxwy} \Bigg[ \matrixel{\Psi^{(0)}_{I}}{\Tilde{E}_{u'u}E_{v'x}E_{wy}}{\Psi^{(0)}_{I}} + \matrixel{\Psi^{(0)}_{I}}{E_{wy}\Tilde{E}_{u'u}E_{v'x}}{\Psi^{(0)}_{I}} + \delta_{v'u'}\Big( \matrixel{\Psi^{(0)}_{I}}{E_{ux}E_{wy}}{\Psi^{(0)}_{I}}\nonumber \\
    &+ \matrixel{\Psi^{(0)}_{I}}{E_{wy}E_{ux}}{\Psi^{(0)}_{I}} \Big) + \delta_{u'w}\matrixel{\Psi^{(0)}_{I}}{\Tilde{E}_{yu}E_{v'x}}{\Psi^{(0)}_{I}} - \delta_{v'y}\matrixel{\Psi^{(0)}_{I}}{\Tilde{E}_{u'u}E_{wx}}{\Psi^{(0)}_{I}}
    \Bigg] \nonumber\\
    & + g_{wxyu} \Bigg[ \matrixel{\Psi^{(0)}_{I}}{E_{wx}\Tilde{E}_{u'y}E_{v'v}}{\Psi^{(0)}_{I}} + \matrixel{\Psi^{(0)}_{I}}{\Tilde{E}_{u'y}E_{v'v}E_{wx}}{\Psi^{(0)}_{I}}  + \delta_{v'u'}\Big( \matrixel{\Psi^{(0)}_{I}}{E_{wx}E_{yv}}{\Psi^{(0)}_{I}} \nonumber\\
    &+ \matrixel{\Psi^{(0)}_{I}}{E_{yv}E_{wx}}{\Psi^{(0)}_{I}} \Big) + \delta_{v'x}\matrixel{\Psi^{(0)}_{I}}{\Tilde{E}_{u'y}E_{wv}}{\Psi^{(0)}_{I}} - \delta_{u'w}\matrixel{\Psi^{(0)}_{I}}{\Tilde{E}_{xy}E_{v'v}}{\Psi^{(0)}_{I}}
    \Bigg] \Bigg\}
\end{align}
The $A_{v'u',uv}$ and $B_{v'u',uv}$ matrices are Hermitian with respect to reference state $\Psi{(0)}$. Furthermore, specific terms arising from the commutator expansion vanish identically as in standard NEVPT2~\cite{Angeli2002}. The contribution arising from  $H_i$ is:
\begin{equation}
    \begin{split}
    \Delta\epsilon_{i,a}^{(0)'}&= \frac{1}{N^{(0)'}_{i,a}} \matrixel{\Psi^{(0)}_{I}}{\left( V^{(0)'}_{i,a} \right) ^\dag \big[H_{i}, V^{(0)'}_{i,a}\big]}{\Psi^{(0)}_{I}}\\
    &= \epsilon_{a} - \epsilon_{i} + \sqrt{\frac{\omega}{2}} \left(d_{aa} - d_{ii}\right) \Bigg\{ 
    \sum_{\Phi} \sum_{m} 2(h^{eff}_{ai})^{2}\left( C_{\Phi,m+1} C_{\Phi,m} \sqrt{m+1} + C_{\Phi,m-1} C_{\Phi,m} \sqrt{m}\right) \\
    & + \sum_{u'v'} \left[2(ai \mid u'v') - (u'i \mid av')\right] h^{eff}_{ai} \gamma_{v'u'} 
    + \sum_{uv} \left[2(ai \mid uv) - (ui \mid av)\right] h^{eff}_{ai} \gamma_{uv} \\
    & + \sum_{\Phi,\Theta} \sum_{m} \sum_{u'v'uv} \Big[ 2(ai \mid u'v')(ai \mid uv) - (u'i \mid av')(ai \mid uv) - (ai \mid u'v')(ui \mid av) \Big] \\
    &\times \left( C_{\Phi,m-1} \matrixel{\Phi}{E_{v'u'}E_{uv}}{\Theta} C_{\Theta,m} \sqrt{m} + C_{\Phi,m+1} \matrixel{\Phi}{E_{v'u'}E_{uv}}{\Theta} C_{\Theta,m} \sqrt{m+1} \right) \\
    & + \sum_{\Phi,\Theta} \sum_{m} \sum_{u'v'uv} (u'i \mid av')(ui \mid av) \Big[ C_{\Phi,m-1} \matrixel{\Phi}{E_{v'v}\Tilde{E}_{u'u} + \delta_{uv}E_{v'u'}}{\Theta} C_{\Theta,m} \sqrt{m} \\
    &+ C_{\Phi,m+1} \matrixel{\Phi}{E_{v'v}\Tilde{E}_{u'u} + \delta_{uv}E_{v'u'}}{\Theta} C_{\Theta,m} \sqrt{m+1} \Big]) \Bigg\} / N^{(0)'}_{i,a}
    \end{split}
\end{equation}
The energy contribution to the second-order energy is given by:
\begin{equation}
    E^{(2)}_{I}(S_{i,a}^{(0)'}) = - \frac{N_{i,a}^{(0)'}}{\Delta\epsilon_{i,a}^{(0)'} + \xi_{i,a}^{(0)'}}
\end{equation}
\subsection{Subspaces  $\mathbf{S_{a}^{(-1)'}}$}
For this subspace, the norm is defined as 
\begin{equation}
    \begin{split}
    N^{(-1)'}_{a} &=\sum_{u'v'w'uvw} ( au' | w'v') ( au | wv ) 
    \matrixel{\Psi^{(0)}_{I}}{E_{v'w'}E_{u'u}E_{wv}}{\Psi^{(0)}_{I}}\\
    &+2 \sum_{u'v'w'u} ( au' | w'v') h^{eff'}_{au} \matrixel{\Psi^{(0)}_{I}}{E_{v'w'}E_{u'u}}{\Psi^{(0)}_{I}}+\sum_{u'u}h^{eff'}_{au'}h^{eff'}_{au}R^{(1)}_{u'u}
    \end{split}
\end{equation}
The contribution arising from  $H_v$ is 
\begin{equation}
    \begin{split}
    \xi^{(-1)'}_{a}&= \frac{1}{N^{(-1)'}_{a}} \matrixel{\Psi^{(0)}_{I}}{\left( V^{(-1)'}_{a} \right) ^\dag \big[H_{v}, V^{(-1)'}_{a}\big]}{\Psi^{(0)}_{I}} \\
    & = \frac{1}{N^{(-1)'}_{a}} \Bigg\{ \sum_{u'v'w'uvw}(au'|w'v')(au|wv)A_{v'w'u',uwv} + \sum_{u'v'w'u} (au'|w'v') h^{eff'}_{au} B_{v'w'u',u} \\
    &+ \sum_{u'uvw} h^{eff'}_{au'} (au|wv) C_{u',uwv} + \sum_{u'u} 
    h^{eff'}_{au'} h^{eff'}_{au} D_{u',u} \Bigg\}
    \end{split}
\end{equation}
We have defined the auxiliary matrices $A_{v'w'u',uwv}$, $B_{v'w'u',u}$ , $C_{u',uwv}$ and $D_{u',u}$ as:
\begin{align}
    A_{v'w'u',uwv} &= \matrixel{\Psi^{(0)}_{I}}{E_{v'w'}E_{u'a}[H_{v},E_{au}E_{wv}]}{\Psi^{(0)}_{I}} \nonumber\\
    &= \sum_{x} \Big[ \overline{h}^{eff}_{xw} \matrixel{\Psi^{(0)}_{I}}{E_{v'w'}E_{u'u}E_{xv}}{\Psi^{(0)}_{I}} - \overline{h}^{eff}_{vx} \matrixel{\Psi^{(0)}_{I}}{E_{v'w'}E_{u'u}E_{wx}}{\Psi^{(0)}_{I}}\nonumber \\
    &- \overline{h}^{eff}_{ux} \matrixel{\Psi^{(0)}_{I}}{E_{v'w'}E_{u'x}E_{wv}}{\Psi^{(0)}_{I}} \Big] + \sum_{xyz} \Big[ 
    g_{xyzw}\matrixel{\Psi^{(0)}_{I}}{E_{v'w'}E_{u'u}E_{xy}E_{zv}}{\Psi^{(0)}_{I}} \nonumber\\
    &- g_{xyvz}\matrixel{\Psi^{(0)}_{I}}{E_{v'w'}E_{u'u}E_{xy}E_{wz}}{\Psi^{(0)}_{I}}  - g_{xyuz}\matrixel{\Psi^{(0)}_{I}}{E_{v'w'}E_{u'z}E_{xy}E_{wv}}{\Psi^{(0)}_{I}} \Big] \nonumber\\
    & + \sum_{xy} \Big[
    g_{xwvy}\matrixel{\Psi^{(0)}_{I}}{E_{v'w'}E_{u'u}E_{xy}}{\Psi^{(0)}_{I}} - \frac{1}{2} g_{xyyw}\matrixel{\Psi^{(0)}_{I}}{E_{v'w'}E_{u'u}E_{xv}}{\Psi^{(0)}_{I}} \nonumber \\
    & - \frac{1}{2} g_{xyvx}\matrixel{\Psi^{(0)}_{I}}{E_{v'w'}E_{u'u}E_{wy}}{\Psi^{(0)}_{I}} + \frac{1}{2} g_{xyux}\matrixel{\Psi^{(0)}_{I}}{E_{v'w'}E_{u'y}E_{wv}}{\Psi^{(0)}_{I}} \Big] \nonumber\\
    & + \sqrt{\frac{\omega}{2}} \sum_{\Phi,\Theta} \sum_{m}  \sum_{x} \Bigg\{ 
    \Big[ C_{\Phi,m-1} \matrixel{\Phi}{E_{v'w'}E_{u'u}E_{xv}}{\Theta} C_{\Theta,m} \sqrt{m} + C_{\Phi,m+1} \matrixel{\Phi}{E_{v'w'}E_{u'u}E_{xv}}{\Theta} C_{\Theta,m} \sqrt{m+1} \Big] d_{xw} \nonumber\\
    & -\Big[ C_{\Phi,m-1} \matrixel{\Phi}{E_{v'w'}E_{u'u}E_{wx}}{\Theta} C_{\Theta,m} \sqrt{m} + C_{\Phi,m+1} \matrixel{\Phi}{E_{v'w'}E_{u'u}E_{wx}}{\Theta} C_{\Theta,m} \sqrt{m+1} \Big] d_{vx} \nonumber\\
    & - \Big[ C_{\Phi,m-1} \matrixel{\Phi}{E_{v'w'}E_{u'x}E_{wv}}{\Theta} C_{\Theta,m} \sqrt{m} + C_{\Phi,m+1} \matrixel{\Phi}{E_{v'w'}E_{u'x}E_{wv}}{\Theta} C_{\Theta,m} \sqrt{m+1} \Big] d_{ux} \Bigg\}  \label{eq:4rdm-1} \\
    \nonumber\\
    B_{v'w'u',u} &= \matrixel{\Psi^{(0)}_{I}}{E_{v'w'}E_{u'a}[\hat{H}_{v},E_{au}]}{\Psi^{(0)}_{I}} \nonumber\\
    &= - \sum_{w} h^{eff'}_{uw} \matrixel{\Psi^{(0)}_{I}}{E_{v'w'}E_{u'w}}{\Psi^{(0)}_{I}} 
    - \sum_{wxy} g_{uwxy} \matrixel{\Psi^{(0)}_{I}}{E_{v'w'}E_{u'w}E_{xy}}{\Psi^{(0)}_{I}} \nonumber\\
    & - \sqrt{\frac{\omega}{2}} \sum_{\Phi,\Theta} \sum_{m} \sum_{w} \Big[ C_{\Phi,m-1} \matrixel{\Phi}{E_{v'w'}E_{u'w}}{\Theta} C_{\Theta,m} \sqrt{m}  + C_{\Phi,m+1} \matrixel{\Phi}{E_{v'w'}E_{u'w}}{\Theta} C_{\Theta,m} \sqrt{m+1} \Big] d_{uw} \\
    \nonumber\\
    C_{u',uwv} &= \matrixel{\Psi^{(0)}_{I}}{E_{u'a}[H_{v},E_{au}E_{wv}]}{\Psi^{(0)}_{I}} \nonumber \\
    &= \sum_{x} \Big[ \overline{h}^{eff}_{xw} \matrixel{\Psi^{(0)}_{I}}{E_{u'u}E_{xv}}{\Psi^{(0)}_{I}} - \overline{h}^{eff}_{vx} \matrixel{\Psi^{(0)}_{I}}{E_{u'u}E_{wx}}{\Psi^{(0)}_{I}} - \overline{h}^{eff}_{ux} \matrixel{\Psi^{(0)}_{I}}{E_{u'x}E_{wv}}{\Psi^{(0)}_{I}} \Big] \nonumber\\
    & + \sum_{xyz} \Big[ g_{xyzw} \matrixel{\Psi^{(0)}_{I}}{E_{u'u}E_{xy}E_{zv}}{\Psi^{(0)}_{I}} - g_{xyvz} \matrixel{\Psi^{(0)}_{I}}{E_{u'u}E_{xy}E_{wz}}{\Psi^{(0)}_{I}}  - g_{xyuz} \matrixel{\Psi^{(0)}_{I}}{E_{u'z}E_{xy}E_{wv}}{\Psi^{(0)}_{I}} \Big] \nonumber\\
    & + \sum_{xy} \Big[ g_{xvwy} \matrixel{\Psi^{(0)}_{I}}{E_{u'u}E_{yx}}{\Psi^{(0)}_{I}} + \frac{1}{2} g_{xyux} \matrixel{\Psi^{(0)}_{I}}{E_{u'y}E_{wv}}{\Psi^{(0)}_{I}} \nonumber\\
    & -  \frac{1}{2} g_{xyyw} \matrixel{\Psi^{(0)}_{I}}{E_{u'u}E_{xv}}{\Psi^{(0)}_{I}} - \frac{1}{2} g_{xyvy} \matrixel{\Psi^{(0)}_{I}}{E_{u'u}E_{wx}}{\Psi^{(0)}_{I}} \Big] \nonumber\\
    & + \sqrt{\frac{\omega}{2}} \sum_{\Phi,\Theta} \sum_{m} \sum_{x} \Bigg\{ 
    \Big[ C_{\Phi,m-1} \matrixel{\Phi}{E_{u'u}E_{xv}}{\Theta} C_{\Theta,m} \sqrt{m} + C_{\Phi,m+1} \matrixel{\Phi}{E_{u'u}E_{xv}}{\Theta} C_{\Theta,m} \sqrt{m+1} \Big] d_{xw} \nonumber\\
    & - \Big[ C_{\Phi,m-1} \matrixel{\Phi}{E_{u'u}E_{wx}}{\Theta} C_{\Theta,m} \sqrt{m} + C_{\Phi,m+1} \matrixel{\Phi}{E_{u'u}E_{wx}}{\Theta} C_{\Theta,m} \sqrt{m+1} \Big] d_{vx} \nonumber\\
    & - \Big[ C_{\Phi,m-1} \matrixel{\Phi}{E_{u'x}E_{wv}}{\Theta} C_{\Theta,m} \sqrt{m} + C_{\Phi,m+1} \matrixel{\Phi}{E_{u'x}E_{wv}}{\Theta} C_{\Theta,m} \sqrt{m+1} \Big] d_{ux} \Bigg\}\\ \nonumber \\
    D_{u'u} &= \matrixel{\Psi^{(0)}_{I}}{E_{u'a}[\hat{H}_{v},E_{au}]}{\Psi^{(0)}_{I}}\nonumber\\
    &= -\sum_{w}h^{eff'}_{uw}R^{(1)}_{u'w}-\sum_{wxy}{g}_{uwxy}\matrixel{\Psi^{(0)}_{I}}{E_{u'w}E_{xy}}{\Psi^{(0)}_{I}} - \sqrt{\frac{\omega}{2}} \sum_w \gamma_{u'w} d_{uw}
\end{align}

For a CAS-CI reference state $\Psi^{(0)}_{I}$, the following relations hold: the $A_{v'w'u',uwv}$ and $D_{u',u}$ matrices are both Hermitian, and $B_{v'w'u',u} = C_{u, u'v'w'}^*$. The contribution arising from  $H_i$ is:
\begin{equation}
    \begin{split}
        \Delta\epsilon^{(-1)'}_{a}&= \frac{1}{N^{(-1)'}_{a}} \matrixel{\Psi^{(0)}_{I}}{\left( V^{(-1)'}_{a} \right) ^\dag \big[H_{i}, V^{(-1)'}_{a}\big]}{\Psi^{(0)}_{I}}\\
        &= \epsilon_{a} + \sqrt{\frac{\omega}{2}} d_{aa} \Bigg\{ 
        \sum_{\Phi,\Theta} \sum_{m} \sum_{u'v'w'uvw} (au' \mid w'v')(au \mid wv) \\
        & \times \Big[ C_{\Phi,m-1} \matrixel{\Phi}{E_{v'w'}E_{u'u}E_{wv}}{\Theta} C_{\Theta,m} \sqrt{m} + C_{\Phi,m+1} \matrixel{\Phi}{E_{v'w'}E_{u'u}E_{wv}}{\Theta} C_{\Theta,m} \sqrt{m+1} \Big] \\
        & + \sum_{\Phi,\Theta} \sum_{m} \sum_{u'v'w'u} (au' \mid w'v') h^{eff'}_{au}\Big[ C_{\Phi,m-1} \matrixel{\Phi}{E_{v'w'}E_{u'u}}{\Theta} C_{\Theta,m} \sqrt{m} + C_{\Phi,m+1} \matrixel{\Phi}{E_{v'w'}E_{u'u}}{\Theta} C_{\Theta,m} \sqrt{m+1} \Big] \\
        & + \sum_{\Phi,\Theta} \sum_{m} \sum_{u'uvw} h^{eff'}_{au'} (au \mid wv)  \Big[ C_{\Phi,m-1} \matrixel{\Phi}{E_{u'u}E_{wv}}{\Theta} C_{\Theta,m} \sqrt{m} + C_{\Phi,m+1} \matrixel{\Phi}{E_{u'u}E_{wv}}{\Theta} C_{\Theta,m} \sqrt{m+1} \Big] \\
        & + \sum_{u'u} h^{eff'}_{au'} h^{eff'}_{au} \gamma_{u'u} \Bigg\} / N^{(-1)'}_{a}
    \end{split}
\end{equation}
The energy contribution to the second-order energy is given by:
\begin{equation}
    E^{(2)}_{I}(S^{(-1)'}_{a}) = - \frac{N^{(-1)'}_{a}}{\Delta\epsilon^{(-1)'}_{a} + \xi^{(-1)'}_{a}}
\end{equation}
\subsection{Subspaces $\mathbf{S_{i}^{(+1)'}}$}
For this subspace, the norm is defined as 
\begin{equation}
    \begin{split}
    N^{(+1)'}_{i} &=\sum_{u'v'w'uvw} ( u'i | w'v') ( ui | wv ) 
    \matrixel{\Psi^{(0)}_{I}}{E_{v'w'}\Tilde{E}_{u'u}E_{wv}}{\Psi^{(0)}_{I}}\\
    &+\sum_{u'uvw} h^{eff}_{u'i} ( ui | wv ) 
    \matrixel{\Psi^{(0)}_{I}}{\Tilde{E}_{u'u}E_{wv}}{\Psi^{(0)}_{I}}\\  
    &+\sum_{u'v'w'u} ( u'i | w'v') h^{eff}_{ui}
    \matrixel{\Psi^{(0)}_{I}}{E_{v'w'}\Tilde{E}_{u'u}}{\Psi^{(0)}_{I}}+\sum_{u'u}h^{eff}_{u'i}h^{eff}_{ui}\Tilde{R}^{(1)}_{u'u}
    \end{split}
\end{equation}
The contribution arising from  $H_v$ is 
\begin{equation}
    \begin{split}
        \xi^{(+1)'}_{i} =& \frac{1}{N^{(+1)'}_{i}} \Bigg\{ \sum_{u'v'w'uvw}(u'i|w'v')
        (ui|wv)\Tilde{A}_{v'w'u',uwv} + \sum_{u'v'w'u}
        (u'i|w'v')h^{eff}_{iu}\Tilde{B}_{v'w'u',u} \\
        &+\sum_{u'uvw}h^{eff}_{iu'}
        (ui|wv)\Tilde{C}_{u',uwv}+\sum_{u'u}h^{eff}_{iu'}
        h^{eff}_{iu}\Tilde{D}_{u'u} \Bigg\}
    \end{split}
\end{equation}
We defined the auxiliary matrices $\Tilde{A}_{v'w'u',uwv}$, $\Tilde{B}_{v'w'u',u}$, $\Tilde{C}_{u',uwv}$ and $\Tilde{D}_{u'u}$
\begin{align}
        \Tilde{A}_{v'w'u',uwv} &= \matrixel{\Psi^{(0)}_{I}}{E_{v'w'}E_{iu'}[H_{v},E_{ui}E_{wv}]}{\Psi^{(0)}_{I}} \nonumber\\
        &= \sum_{x} \Big[ \overline{h}^{eff}_{xw} \matrixel{\Psi^{(0)}_{I}}{E_{v'w'}\Tilde{E}_{u'u}E_{xv}}{\Psi^{(0)}_{I}} + \overline{h}^{eff}_{xu} \matrixel{\Psi^{(0)}_{I}}{E_{v'w'}\Tilde{E}_{u'x}E_{wv}}{\Psi^{(0)}_{I}} \nonumber \\ 
        &- \overline{h}^{eff}_{vx} \matrixel{\Psi^{(0)}_{I}}{E_{v'w'}\Tilde{E}_{u'u}E_{wx}}{\Psi^{(0)}_{I}} \Big]  + \sum_{xyz} \Big[ 
        g_{xyzw}\matrixel{\Psi^{(0)}_{I}}{E_{v'w'}\Tilde{E}_{u'u}E_{xy}E_{zv}}{\Psi^{(0)}_{I}} \nonumber \\
        &- g_{xyvz}\matrixel{\Psi^{(0)}_{I}}{E_{v'w'}\Tilde{E}_{u'u}E_{xy}E_{wz}}{\Psi^{(0)}_{I}}
        + g_{xyuz}\matrixel{\Psi^{(0)}_{I}}{E_{v'w'}\Tilde{E}_{u'z}E_{xy}E_{wv}}{\Psi^{(0)}_{I}} \Big] \nonumber \\
        &+ \sum_{xy} \Big[ 
        g_{xwvy}\matrixel{\Psi^{(0)}_{I}}{E_{v'w'}\Tilde{E}_{u'u}E_{xy}}{\Psi^{(0)}_{I}} - \frac{1}{2} g_{xyyw}\matrixel{\Psi^{(0)}_{I}}{E_{v'w'}\Tilde{E}_{u'u}E_{xv}}{\Psi^{(0)}_{I}} \nonumber\\
        & - \frac{1}{2} g_{xyvx}\matrixel{\Psi^{(0)}_{I}}{E_{v'w'}\Tilde{E}_{u'u}E_{wy}}{\Psi^{(0)}_{I}} + \frac{1}{2} g_{xyux}\matrixel{\Psi^{(0)}_{I}}{E_{v'w'}\Tilde{E}_{u'y}E_{wv}}{\Psi^{(0)}_{I}} \Big] \nonumber\\
        & + \sqrt{\frac{\omega}{2}} \sum_{\Phi,\Theta} \sum_{m} \sum_{x} \Bigg\{ 
         \Big[ C_{\Phi,m-1} \matrixel{\Phi}{E_{v'w'}\Tilde{E}_{u'u}E_{xv}}{\Theta} C_{\Theta,m} \sqrt{m} + C_{\Phi,m+1} \matrixel{\Phi}{E_{v'w'}\Tilde{E}_{u'u}E_{xv}}{\Theta} C_{\Theta,m} \sqrt{m+1} \Big] d_{xw} \nonumber\\
        & - \Big[ C_{\Phi,m-1} \matrixel{\Phi}{E_{v'w'}\Tilde{E}_{u'u}E_{wx}}{\Theta} C_{\Theta,m} \sqrt{m} + C_{\Phi,m+1} \matrixel{\Phi}{E_{v'w'}\Tilde{E}_{u'u}E_{wx}}{\Theta} C_{\Theta,m} \sqrt{m+1} \Big] d_{vx} \nonumber\\
        & + \Big[ C_{\Phi,m-1} \matrixel{\Phi}{E_{v'w'}\Tilde{E}_{u'x}E_{wv}}{\Theta} C_{\Theta,m} \sqrt{m} + C_{\Phi,m+1} \matrixel{\Phi}{E_{v'w'}\Tilde{E}_{u'x}E_{wv}}{\Theta} C_{\Theta,m} \sqrt{m+1} \Big] d_{xu} \Bigg\} \label{eq:4rdm-2}\\ \nonumber \\
        \Tilde{B}_{v'w'u',u} &= \matrixel{\Psi^{(0)}_{I}}{E_{v'w'}E_{iu'}[H_{v},E_{ui}]}{\Psi^{(0)}_{I}} \nonumber \\
        &= \sum_{w} h^{eff'}_{wu} \matrixel{\Psi^{(0)}_{I}}{E_{v'w'}\Tilde{E}_{u'w}}{\Psi^{(0)}_{I}} 
        + \sum_{wxy} g_{uwxy} \matrixel{\Psi^{(0)}_{I}}{E_{v'w'}\Tilde{E}_{u'w}E_{xy}}{\Psi^{(0)}_{I}}\nonumber  \\
        & + \sqrt{\frac{\omega}{2}} \sum_{\Phi,\Theta} \sum_{m}  \sum_{w} \Big[ C_{\Phi,m-1} \matrixel{\Phi}{E_{v'w'}\Tilde{E}_{u'w}}{\Theta} C_{\Theta,m} \sqrt{m}  + C_{\Phi,m+1} \matrixel{\Phi}{E_{v'w'}\Tilde{E}_{u'w}}{\Theta} C_{\Theta,m} \sqrt{m+1} \Big] d_{wu} \\ \nonumber \\
        \Tilde{C}_{u',uwv} &= \matrixel{\Psi^{(0)}_{I}}{E_{iu'}[H_{v},E_{ui}E_{wv}]}{\Psi^{(0)}_{I}} \nonumber\\
        &= \sum_{x} \Big[ \overline{h}^{\text{eff}}_{xw} \matrixel{\Psi^{(0)}_{I}}{\Tilde{E}_{u'u}E_{xv}}{\Psi^{(0)}_{I}} + \overline{h}^{\text{eff}}_{xu} \matrixel{\Psi^{(0)}_{I}}{\Tilde{E}_{u'x}E_{wv}}{\Psi^{(0)}_{I}}  - \overline{h}^{\text{eff}}_{vx} \matrixel{\Psi^{(0)}_{I}}{\Tilde{E}_{u'u}E_{wx}}{\Psi^{(0)}_{I}} \Big] \nonumber\\
        & + \sum_{xyz} \Big[ g_{xyzw} \matrixel{\Psi^{(0)}_{I}}{\Tilde{E}_{u'u}E_{xy}E_{zv}}{\Psi^{(0)}_{I}} - g_{xyvz} \matrixel{\Psi^{(0)}_{I}}{\Tilde{E}_{u'u}E_{xy}E_{wz}}{\Psi^{(0)}_{I}} \nonumber\\
        & + g_{xyuz} \matrixel{\Psi^{(0)}_{I}}{\Tilde{E}_{u'z}E_{xy}E_{wv}}{\Psi^{(0)}_{I}} \Big] \nonumber\\
        & + \sum_{xy} \Big[ g_{xwvy} \matrixel{\Psi^{(0)}_{I}}{\Tilde{E}_{u'u}E_{xy}}{\Psi^{(0)}_{I}} - \frac{1}{2} g_{xyyw} \matrixel{\Psi^{(0)}_{I}}{\Tilde{E}_{u'u}E_{xv}}{\Psi^{(0)}_{I}} \nonumber\\
        & - \frac{1}{2} g_{xyvx} \matrixel{\Psi^{(0)}_{I}}{\Tilde{E}_{u'u}E_{wy}}{\Psi^{(0)}_{I}} + \frac{1}{2} g_{xyux} \matrixel{\Psi^{(0)}_{I}}{\Tilde{E}_{u'y}E_{wv}}{\Psi^{(0)}_{I}} \Big] \nonumber\\
        & + \sqrt{\frac{\omega}{2}} \sum_{\Phi,\Theta} \sum_{m} \sum_x\Bigg\{ 
        \Big[ C_{\Phi,m-1} \matrixel{\Phi}{\Tilde{E}_{u'u}E_{xv}}{\Theta} C_{\Theta,m} \sqrt{m} + C_{\Phi,m+1} \matrixel{\Phi}{\Tilde{E}_{u'u}E_{xv}}{\Theta} C_{\Theta,m} \sqrt{m+1} \Big] d_{xw} \nonumber \\
        & - \Big[ C_{\Phi,m-1} \matrixel{\Phi}{\Tilde{E}_{u'u}E_{wx}}{\Theta} C_{\Theta,m} \sqrt{m} + C_{\Phi,m+1} \matrixel{\Phi}{\Tilde{E}_{u'u}E_{wx}}{\Theta} C_{\Theta,m} \sqrt{m+1} \Big] d_{vx} \nonumber \\
        & + \Big[ C_{\Phi,m-1} \matrixel{\Phi}{\Tilde{E}_{u'x}E_{wv}}{\Theta} C_{\Theta,m} \sqrt{m} + C_{\Phi,m+1} \matrixel{\Phi}{\Tilde{E}_{u'x}E_{wv}}{\Theta} C_{\Theta,m} \sqrt{m+1} \Big] d_{xu} \Bigg\} \\ \nonumber \\
        \Tilde{D}_{u',u}&=\matrixel{\Psi^{(0)}_{I}}{E_{iu'}[\hat{H}_{v},E_{ui}]}{\Psi^{(0)}_{I}}\nonumber\\
        &=\sum_{w}{h}^{eff}_{wu} \Tilde{R}^{(1)}_{u'w}    +\sum_{wxy}{g}_{uwxy}\matrixel{\Psi^{(0)}_{I}}{\Tilde{E}_{u'w}E_{xy}}{\Psi^{(0)}_{I}} + \sqrt{\frac{\omega}{2}}\sum_w \Tilde{\gamma}_{u'w}d_{wu} 
\end{align}
For a CAS-CI reference state $\Psi^{(0)}_{I}$, the following relations hold: the $\tilde{A}_{v'w'u',uwv}$ and $\tilde{D}_{u',u}$ matrices are both Hermitian, and $\tilde{B}_{v'w'u',u} = \tilde{C}_{u',uwv}^*$. The contribution arising from $H_i$ is:
\begin{equation}
    \begin{split}
        \Delta\epsilon^{(+1)'}_{i} &= \frac{1}{N^{(+1)'}_{i}} \matrixel{\Psi^{(0)}_{I}}{\left( V^{(+1)'}_{i} \right) ^\dag \big[H_{i}, V^{(+1)'}_{i}\big]}{\Psi^{(0)}_{I}}\\
        &= - \epsilon_{i} - \sqrt{\frac{\omega}{2}} d_{ii} \Bigg\{ 
        \sum_{\Phi,\Theta} \sum_{m} \sum_{u'v'w'uvw} (u'i \mid w'v')(ui \mid wv) \\
        & \times \left( C_{\Phi,m-1} \matrixel{\Phi}{E_{v'w'}\Tilde{E}_{u'u}E_{wv}}{\Theta} C_{\Theta,m} \sqrt{m} + C_{\Phi,m+1} \matrixel{\Phi}{E_{v'w'}\Tilde{E}_{u'u}E_{wv}}{\Theta} C_{\Theta,m} \sqrt{m+1} \right) \\
        & + \sum_{\Phi,\Theta} \sum_{m} \sum_{u'uvw} h^{eff}_{u'i} (ui \mid wv) \\
        & \times \left( C_{\Phi,m-1} \matrixel{\Phi}{\Tilde{E}_{u'u}E_{wv}}{\Theta} C_{\Theta,m} \sqrt{m} + C_{\Phi,m+1} \matrixel{\Phi}{\Tilde{E}_{u'u}E_{wv}}{\Theta} C_{\Theta,m} \sqrt{m+1} \right) \\
        & + \sum_{\Phi,\Theta} \sum_{m} \sum_{u'v'w'u} (u'i \mid w'v') h^{eff}_{ui} \\
        & \times \left( C_{\Phi,m-1} \matrixel{\Phi}{E_{v'w'}\Tilde{E}_{u'u}}{\Theta} C_{\Theta,m} \sqrt{m} + C_{\Phi,m+1} \matrixel{\Phi}{E_{v'w'}\Tilde{E}_{u'u}}{\Theta} C_{\Theta,m} \sqrt{m+1} \right) \\
        & + \sum_{u'u} h^{eff}_{u'i}h^{eff}_{ui} \Tilde{\gamma}_{u'u} \Bigg\} / N^{(+1)'}_{i}
    \end{split}
\end{equation}
The energy contribution to the second-order energy is given by:
\begin{equation}
    E^{(2)}_{I}(S^{(+1)'}_{i}) = - \frac{N^{(+1)'}_{i}}{\Delta\epsilon^{(+1)'}_{i} + \xi^{(+1)'}_{i}}
\end{equation}
\section{Mixed-plus perturbers}\label{AppendixB}
Unlike standard electronic perturbers, mixed-excitation terms do not commute with the photonic Hamiltonian, preventing the usual cancellation of some terms and introducing additional contributions.
\subsection{Subspaces $\mathbf{S_{i,a,+}^{(0)'}}$}
The norm, for this subspace, is defined as
\begin{equation}
    N_{i,a,+}^{(0)'} = \omega  d_{ai}^{2} \sum_{\Phi} \sum_{m} |C_{\Phi,m}|^2 \left(m + 1\right) 
\end{equation}
The contribution arising from  $H_i$ is 
\begin{equation}
    \begin{split}
    \Delta\epsilon_{i,a,+}^{(0)'}&= \epsilon_a  - \epsilon_i + 2 \left(\frac{\omega}{2}\right)^{\frac{3}{2}} d_{ai}^{2}\\
    &\times \Bigg\{d_{aa} \Big[ \sum_{\Phi} \sum_{m} C_{\Phi,m+1} C_{\Phi,m} \sqrt{m+1} (m + 2) \\
        &+\sum_{\Phi} \sum_{m} C_{\Phi,m-1} C_{\Phi,m} \sqrt{m} (m + 1)\Big] \\    
        & - d_{ii} \Big[ \sum_{\Phi} \sum_{m} C_{\Phi,m+1} C_{\Phi,m} \sqrt{m+1} (m + 2) \\
        & + \sum_{\Phi} \sum_{m} C_{\Phi,m-1} C_{\Phi,m} \sqrt{m} (m + 1) \Big] \\
        &+ 2\sum_j d_{jj} \sum_{\Phi} \sum_{m} C_{\Phi,m-1} C_{\Phi,m} \sqrt{m} \Bigg\}/ N_{i,a,+}^{(0)'}
    \end{split}
\end{equation}
In this case the contribution from $H_v$ is non-zero and is given by: 
\begin{equation}
    \xi^{(0)'}_{i,a,+} = \frac{1}{N^{(0)'}_{i,a,+}}  \matrixel{\Psi^{(0)}_{I}}{bE_{ia} [H_{v}, E_{ai}b^{\dag}]}{\Psi^{(0)}_{I}} 
\end{equation}
where
\begin{equation}
    \begin{split}
    &\matrixel{\Psi^{(0)}_{I}}{bE_{ia} [H_{v}, E_{ai}b^{\dag}]}{\Psi^{(0)}_{I}} = 2 \left(\frac{\omega}{2}\right)^{\frac{3}{2}} (d_{ai})^2 \\
    &\times \sum_{uv} d_{uv} \sum_{\Phi,\Theta} \sum_{m} C_{\Phi,m-1} \matrixel{\Phi}{E_{uv}}{\Theta} C_{\Theta,m} \sqrt{m}
    \end{split}
\end{equation}
The contribution arising from  $H_{ph}$ is 
\begin{equation}
    \frac{1}{N_{i,a,+}^{(0)'}} \langle \Psi^{(0)}_{I} \rvert ({V}_{i,a,+}^{(0)'})^{\dagger} [{H}_{ph}, {V}_{i,a,+}^{(0)'}] \lvert \Psi^{(0)}_{I} \rangle = \omega
\end{equation}
The contribution to the second-order energy is given  by:
\begin{equation}
    E^{(2)}_{I} (S_{i,a,+}^{(0)'})= - \frac{N_{i,a,+}^{(0)'}}{\Delta\epsilon_{i,a,+}^{(0)'} + \xi^{(0)'}_{i,a,+} + \omega}
\end{equation}
\subsection{Subspaces $\mathbf{S_{i,+}^{(+1)'}}$}
The evaluation of energy contribution of for $S_{i,+}^{(+1)'}$ is very similar to the previous one. In this case, the norm is defined as
\begin{equation}
    N_{i,+}^{(+1)'} = \frac{\omega}{2} \sum_{\Phi,\Theta}\sum_{m} \sum_{u'u} d_{u'i} d_{ui} C_{\Phi,m} \matrixel{\Phi}{\Tilde{E}_{u'u}}{\Theta} C_{\Theta,m} (m+1)
\end{equation}
The contribution arising from  $H_i$ is 
\begin{equation}
    \begin{split}
    &\Delta\epsilon_{i,+}^{(+1)'}= - \epsilon_i -  \left(\frac{\omega}{2}\right)^{\frac{3}{2}} \sum_{u'u}d_{ui} d_{u'i} \\
    & \times \Bigg\{ d_{ii} \Big[ \sum_{\Phi,\Theta} \sum_{m} C_{\Phi,m+1} \matrixel{\Phi}{\Tilde{E}_{u'u}}{\Theta} C_{\Theta,m} \sqrt{m+1} (m + 2) \\
        &+\sum_{\Phi,\Theta} \sum_{m} C_{\Phi,m-1} \matrixel{\Phi}{\Tilde{E}_{u'u}}{\Theta} C_{\Theta,m} \sqrt{m} (m + 1)\Big] \\
        &- 2 \sum_j d_{jj} \sum_{\Phi,\Theta} \sum_{m} C_{\Phi, m-1}\matrixel{\Phi}{\Tilde{E}_{u'u}}{\Theta} C_{\Theta,m} \sqrt{m} \Bigg\} / N_{i,+}^{(+1)'}
    \end{split}
\end{equation}
The active contribution is 
\begin{equation}
    \xi_{i,+}^{(+1)'} = \frac{1}{N_{i,+}^{(+1)'}} \langle \Psi^{(0)}_{I} \rvert ({V}_{i,+}^{(+1)'})^{\dagger} [{H}_{v}, {V}_{i,+}^{(+1)'}] \lvert \Psi^{(0)}_{I} \rangle 
\end{equation}
where
\begin{equation}
    \begin{split}
     &\langle \Psi^{(0)}_{I} \rvert ({V}_{i,+}^{(+1)'})^{\dagger} [{H}_{v}, {V}_{i,+}^{(+1)'}] \lvert \Psi^{(0)}_{I} \rangle = \\
     &\frac{\omega}{2}  \sum_{u'u} d_{u'i} d_{ui} \Bigg\{ \sum_{v} h^{eff}_{vu} \sum_{\Phi,\Theta} \sum_{m} C_{\Phi,m} \matrixel{\Phi}{\Tilde{E}_{u'v}}{\Theta} C_{\Theta,m} (m +1) \\
     &+ \sum_{wxy} g_{uwxy} \sum_{\Phi,\Theta} \sum_{m} C_{\Phi,m} \matrixel{\Phi}{\Tilde{E}_{u'w}E_{xy}}{\Theta} C_{\Theta,m} (m + 1) \\
     & + \sqrt{\frac{\omega}{2}}\sum_v d_{vu} \Big[ \sum_{\Phi,\Theta} \sum_{m} C_{\Phi,m+1} \matrixel{\Phi}{\Tilde{E}_{u'v}}{\Theta} C_{\Theta,m} \sqrt{m+1} (m + 2) \\
     & +\sum_{\Phi,\Theta} \sum_{m} C_{\Phi,m-1} \matrixel{\Phi}{\Tilde{E}_{u'v}}{\Theta} C_{\Theta,m} \sqrt{m} (m + 1)\Big] \\
     & + \sqrt{\frac{\omega}{2}} \sum_{vw} d_{vw} \sum_{\Phi,\Theta} \sum_{m} C_{\Phi,m-1} \matrixel{\Phi}{\Tilde{E}_{u'u}E_{vw}}{\Theta} C_{\Theta,m} \sqrt{m}
     \Bigg\}
     \end{split}
\end{equation}
The contribution arising from  $H_{ph}$, due to normalization, is equal to the rest of the previous mixed class perturbers.
\begin{equation}
    \frac{1}{N_{i,+}^{(+1)'}} \langle \Psi^{(0)}_{I} \rvert ({V}_{i,+}^{(+1)'})^{\dagger} [{H}_{ph}, {V}_{i,+}^{(+1)'}] \lvert \Psi^{(0)}_{I} \rangle = \omega
\end{equation}
The contribution to the second-order energy is given  by:
\begin{equation}
    E^{(2)}_{I}(S_{i,+}^{(+1)'})= - \frac{N_{i,+}^{(+1)'}}{\Delta\epsilon_{i,+}^{(+1)'} + \xi_{i,+}^{(+1)'} + \omega}
\end{equation}
\section{Mixed-minus perturbers}\label{AppendixC}
The theoretical treatment for this class of perturbers closely parallels that of the $\ket{K_+}$ manifold.
\subsection{Subspaces $\mathbf{S_{i,a,-}^{(0)'}}$}
The norm for this subspace is defined as
\begin{equation}
    N_{i,a,-}^{(0)'} = \omega  d_{ai}^{2} \sum_{\Phi} \sum_{m} |C_{\Phi,m}|^2 m 
\end{equation}
The contribution arising from  $H_i$ is 
\begin{equation}
    \begin{split}
    \Delta\epsilon_{i,a,-}^{(0)'}&= \epsilon_a  - \epsilon_i + 2 \left(\frac{\omega}{2}\right)^{\frac{3}{2}} d_{ai}^{2}\\
    &\times \Bigg\{d_{aa} \Big[ \sum_{\Phi} \sum_{m} C_{\Phi,m+1} C_{\Phi,m} m \sqrt{m+1}  \\
        &+\sum_{\Phi} \sum_{m} C_{\Phi,m-1} C_{\Phi,m} \sqrt{m} (m - 1)\Big] \\    
        & - d_{ii} \Big[ \sum_{\Phi} \sum_{m} C_{\Phi,m+1} C_{\Phi,m} m \sqrt{m+1} \\
        & + \sum_{\Phi} \sum_{m} C_{\Phi,m-1} C_{\Phi,m} \sqrt{m} (m - 1)\Big] \\
        &- 2\sum_j d_{jj} \sum_{\Phi} \sum_{m} C_{\Phi,m+1} C_{\Phi,m} \sqrt{m + 1}\Bigg\}/ N_{i,a,-}^{(0)'}
    \end{split}
\end{equation}
In this case the contribution from $H_v$ is non-zero and is given by: 
\begin{equation}
    \xi^{(0)'}_{i,a,-} = \frac{1}{N^{(0)'}_{i,a,-}}  \matrixel{\Psi^{(0)}_{I}}{b^{\dag} E_{ia} [H_{v}, E_{ai}b]}{\Psi^{(0)}_{I}} 
\end{equation}
where
\begin{equation}
    \begin{split}
    &\matrixel{\Psi^{(0)}_{I}}{b^{\dag} E_{ia} [H_{v}, E_{ai}b]}{\Psi^{(0)}_{I}}  = - 2 \left(\frac{\omega}{2}\right)^{\frac{3}{2}} (d_{ai})^2 \\
    &\times \sum_{uv} d_{uv} \sum_{\Phi,\Theta} \sum_{m} C_{\Phi,m+1} \matrixel{\Phi}{E_{uv}}{\Theta} C_{\Theta,m} \sqrt{m+1}
    \end{split}
\end{equation}
The contribution arising from  $H_{ph}$ is 
\begin{equation}
    \frac{1}{N_{i,a,-}^{(0)'}} \langle \Psi^{(0)}_{I} \rvert ({V}_{i,a,-}^{(0)'})^{\dagger} [{H}_{ph}, {V}_{i,a,-}^{(0)'}] \lvert \Psi^{(0)}_{I} \rangle = - \omega
\end{equation}
The contribution to the second-order energy is given  by:
\begin{equation}
    E^{(2)}_{I} (S_{i,a,-}^{(0)'})= - \frac{N_{i,a,-}^{(0)'}}{\Delta\epsilon_{i,a,-}^{(0)'} + \xi^{(0)'}_{i,a,-} - \omega}
\end{equation}
\subsection{Subspaces $\mathbf{S_{a,-}^{(-1)'}}$}
For this subspace the norm is defined as 
\begin{equation}
    N_{a,-}^{(-1)'}
    =\frac{\omega}{2}  \sum_{u'u} d_{au} d_{au'}\sum_{\Phi,\Theta}\sum_{m} C_{\Phi,m} \matrixel{\Phi}{E_{u'u}}{\Theta} C_{\Theta,m} m
\end{equation}
The contribution arising from  $H_i$ is 
\begin{equation}
    \begin{split}
    &\Delta\epsilon_{a,-}^{(-1)'} = \epsilon_a  + \left( \frac{\omega}{2} \right)^{\frac{3}{2}} \sum_{u'u} d_{au'} d_{au}   \\
    & \times \Bigg\{ d_{aa} \Big[ \sum_{\Phi,\Theta} \sum_{m} C_{\Phi,m+1} \matrixel{\Phi}{E_{u'u}}{\Theta} C_{\Theta,m} m\sqrt{m+1} \\
    &+\sum_{\Phi,\Theta} \sum_{m} C_{\Phi,m-1} \matrixel{\Phi}{E_{u'u}}{\Theta} C_{\Theta,m} \sqrt{m} (m - 1)\Big]  \\
    &- 2\sum_i d_{ii} \sum_{\Phi,\Theta} \sum_{m} C_{\Phi,m+1} \matrixel{\Phi}{E_{u'u}}{\Theta} C_{\Theta,m} \sqrt{m + 1}
        \Bigg\} / N_{a,-}^{(-1)'}
    \end{split}
\end{equation}
The contribution from the valence term of the QED-Dyall Hamiltonian is
\begin{equation}
    \begin{split}
    \xi_{a,-}^{(-1)'} = \frac{1}{N_{a,-}^{(-1)'}} \langle \Psi^{(0)}_{I} | ({V}_{a,-}^{(-1)'})^{\dagger} [{H}_v, {V}_{a,-}^{(-1)'}] | \Psi^{(0)}_{I} \rangle 
    \end{split}
\end{equation}
where
\begin{equation}
    \begin{split}
    &\langle \Psi^{(0)}_{I} | ({V}_{a,-}^{(-1)'})^{\dagger} [{H}_v, {V}_{a,-}^{(-1)'}] | \Psi^{(0)}_{I} \rangle \\
    &= \frac{\omega}{2} \sum_{u'u} d_{au'} d_{au} \Big\{ - \sum_{w} h^{eff}_{uw} \\
    & \times \sum_{\Phi,\Theta} \sum_{m} C_{\Phi,m} \matrixel{\Phi}{E_{u'w}}{\Theta} C_{\Theta,m} m\\
    &- \sum_{wxy} g_{uwxy} \sum_{\Phi,\Theta} \sum_{m} C_{\Phi,m} \matrixel{\Phi}{E_{u'w}E_{xy}}{\Theta} C_{\Theta,m} m \\
    &- \sqrt{\frac{\omega}{2}} \sum_w d_{uw} \Big[ \sum_{\Phi,\Theta} \sum_{m} C_{\Phi,m+1} \matrixel{\Phi}{E_{u'w}}{\Theta} C_{\Theta,m} m \sqrt{m+1} \\
        &+\sum_{\Phi,\Theta} \sum_{m} C_{\Phi,m-1} \matrixel{\Phi}{E_{u'w}}{\Theta} C_{\Theta,m} \sqrt{m} (m - 1) \Big] \\
    &- \sqrt{\frac{\omega}{2}}\sum_{vw} d_{vw} \sum_{\Phi,\Theta} \sum_{m}  C_{\Phi,m+1} \matrixel{\Phi}{E_{u'u}E_{vw}}{\Theta} C_{\Theta,m} \sqrt{m + 1} \Big\}
    \end{split}
\end{equation}
The photonic contribution is
\begin{equation}
    \begin{split}
    & \frac{1}{N_{a,-}^{(-1)'}} \langle \Psi^{(0)}_{I} \rvert ({V}_{a,-}^{(-1)'})^{\dagger} [{H}_{ph}, {V}_{a,-}^{(-1)'}] \lvert \Psi^{(0)}_{I} \rangle = - \omega
    \end{split}
\end{equation}
Therefore, the contribution to the second-order energy is:
\begin{equation}
    E^{(2)}_{I}(S_{a,-}^{(-1)'})=- \frac{N_{a,-}^{(-1)'}}{\Delta\epsilon_{a,-}^{(-1)'} + \xi_{a,-}^{(-1)'} - \omega} 
\end{equation}
\subsection{Subspaces $\mathbf{S_{i,-}^{(+1)'}}$}
The evaluation of energy contribution of for $S_{i,-}^{(+1)'}$ is very similar to the previous one. In this case, the norm is defined as
\begin{equation}
    N_{i,-}^{(+1)'} = \frac{\omega}{2} \sum_{\Phi,\Theta}\sum_{m} \sum_{u'u} d_{u'i} d_{ui} C_{\Phi,m} \matrixel{\Phi}{\Tilde{E}_{u'u}}{\Theta} C_{\Theta,m} m
\end{equation}
The contribution arising from  $H_i$ is 
    \begin{equation}
        \begin{split}
        &\Delta\epsilon_{i,-}^{(+1)'}= - \epsilon_i - \left(\frac{\omega}{2}\right)^{\frac{3}{2}} \sum_{u'u}d_{ui} d_{u'i} \\
        & \times \Bigg\{ d_{ii} \Big[ \sum_{\Phi,\Theta} \sum_{m} C_{\Phi,m+1} \matrixel{\Phi}{\Tilde{E}_{u'u}}{\Theta} C_{\Theta,m} m \sqrt{m+1} \\
            &+\sum_{\Phi,\Theta} \sum_{m} C_{\Phi,m-1} \matrixel{\Phi}{\Tilde{E}_{u'u}}{\Theta} C_{\Theta,m} \sqrt{m} (m - 1)\Big] \\
            &+ 2 \sum_j d_{jj} \sum_{\Phi,\Theta} \sum_{m} C_{\Phi, m+1} \matrixel{\Phi}{\Tilde{E}_{u'u}}{\Theta} C_{\Theta,m} \sqrt{m + 1} \Bigg\} / N_{i,-}^{(+1)'}
        \end{split}
    \end{equation}
The active contribution is 
\begin{equation}
    \xi_{i,-}^{(+1)'} = \frac{1}{N_{i,-}^{(+1)'}} \langle \Psi^{(0)}_{I} \rvert ({V}_{i,-}^{(+1)'})^{\dagger} [{H}_{v}, {V}_{i,-}^{(+1)'}] \lvert \Psi^{(0)}_{I} \rangle 
\end{equation}
where
\begin{equation}
    \begin{split}
     &\langle \Psi^{(0)}_{I} \rvert ({V}_{i,-}^{(+1)'})^{\dagger} [{H}_{v}, {V}_{i,-}^{(+1)'}] \lvert \Psi^{(0)}_{I} \rangle = \\
     &\frac{\omega}{2}  \sum_{u'u} d_{u'i} d_{ui} \Bigg\{\sum_{v} h^{eff}_{vu} \sum_{\Phi,\Theta} \sum_{m} C_{\Phi,m} \matrixel{\Phi}{\Tilde{E}_{u'v}}{\Theta} C_{\Theta,m} m \\
     &+ \sum_{wxy} g_{uwxy} \sum_{\Phi,\Theta} \sum_{m} C_{\Phi,m} \matrixel{\Phi}{\Tilde{E}_{u'w}E_{xy}}{\Theta} C_{\Theta,m} m \\
     & + \sqrt{\frac{\omega}{2}}\sum_v d_{vu} \Big[ \sum_{\Phi,\Theta} \sum_{m} C_{\Phi,m+1} \matrixel{\Phi}{\Tilde{E}_{u'v}}{\Theta} C_{\Theta,m} m \sqrt{m+1} \\
        &+\sum_{\Phi,\Theta} \sum_{m} C_{\Phi,m-1} \matrixel{\Phi}{\Tilde{E}_{u'v}}{\Theta} C_{\Theta,m} \sqrt{m} (m - 1)\Big] \\
     & - \sqrt{\frac{\omega}{2}} \sum_{vw} d_{vw} \sum_{\Phi,\Theta} \sum_{m} C_{\Phi,m+1} \matrixel{\Phi}{\Tilde{E}_{u'u}E_{vw}}{\Theta} C_{\Theta,m} \sqrt{m + 1}
     \Bigg\}
     \end{split}
\end{equation}
The contribution arising from  $H_{ph}$, due to normalization, is equal to the rest of the previous mixed class perturbers.
\begin{equation}
    \frac{1}{N_{i,-}^{(+1)'}} \langle \Psi^{(0)}_{I} \rvert ({V}_{i,-}^{(+1)'})^{\dagger} [{H}_{ph}, {V}_{i,-}^{(+1)'}] \lvert \Psi^{(0)}_{I} \rangle = - \omega
\end{equation}
The contribution to the second-order energy is given  by:
\begin{equation}
    E^{(2)}_{I}(S_{i,-}^{(+1)'})= - \frac{N_{i,-}^{(+1)'}}{\Delta\epsilon_{i,-}^{(+1)'} + \xi_{i,-}^{(+1)'} - \omega}
\end{equation}

\end{widetext}
%
%

\clearpage
\onecolumngrid
\foreach \p in {1,...,5}{%
  \clearpage
  \thispagestyle{empty}%
  \null
  \vspace*{-\voffset}\vspace*{-1in}\vspace*{-\topmargin}\vspace*{-\headheight}\vspace*{-\headsep}%
  \hspace*{-\hoffset}\hspace*{-1in}\hspace*{-\oddsidemargin}%
  \makebox[0pt][l]{\includegraphics[page=\p,width=\paperwidth,height=\paperheight]{SI_QED_NEVPT2_Paper.pdf}}%
}
\end{document}